\documentclass[english,twocolumn,prx,amssymb,amsmath,superscriptaddress,showpacs]{revtex4-1}
\usepackage[latin9]{inputenc}
\usepackage{graphicx,color}
\usepackage[bookmarks]{hyperref}
\usepackage{babel}
\usepackage{braket}

\newcommand{\be}{\begin{equation}}
\newcommand{\ee}{\end{equation}}
\newcommand{\bea}{\begin{eqnarray}}
\newcommand{\eea}{\end{eqnarray}}

\newcommand{\ue}{\ensuremath{{\rm{e}}}}
\newcommand{\ud}{\ensuremath{{\rm{d}}}}

\newcommand{\obs}{\hat{\cal O}}
\newcommand{\hb}{\hat{b}} 

\begin{document}

\title{Measuring multipartite entanglement via dynamic susceptibilities}

\author{Philipp Hauke}
	\email{philipp.hauke@uibk.ac.at}    
	\affiliation{Institute for Theoretical Physics, University of Innsbruck, 6020 Innsbruck, Austria}
	\affiliation{Institute for Quantum Optics and Quantum Information of the Austrian Academy of Sciences, 6020 Innsbruck, Austria}
\author{Markus Heyl}
	\affiliation{Institute for Theoretical Physics, University of Innsbruck, 6020 Innsbruck, Austria}
	\affiliation{Institute for Quantum Optics and Quantum Information of the Austrian Academy of Sciences, 6020 Innsbruck, Austria}
	\affiliation{	Physik Department, Technische Universit\"at M\"unchen, 85747 Garching, Germany}	
\author{Luca Tagliacozzo}
	\affiliation{ICFO-Institut de Ciencies Fotoniques, Mediterranean Technology Park, 08860 Castelldefels (Barcelona), Spain}
	\affiliation{Department of Physics and Scottish Universities Physics Alliance University of Strathclyde, Glasgow G4 0NG, Scotland, UK}
\author{Peter Zoller}
	\affiliation{Institute for Theoretical Physics, University of Innsbruck, 6020 Innsbruck, Austria}
	\affiliation{Institute for Quantum Optics and Quantum Information of the Austrian Academy of Sciences, 6020 Innsbruck, Austria}

\begin{abstract}
Entanglement plays a central role in our understanding of quantum many body physics, and is fundamental in characterising quantum phases and quantum phase transitions. Developing protocols to detect and quantify entanglement of many-particle quantum states is thus a key challenge for present experiments. Here, we show that the quantum Fisher information, representing a witness for genuinely multipartite entanglement, becomes measurable for thermal ensembles via the dynamic susceptibility, i.e., with resources readily available in present cold atomic gas and condensed-matter experiments. 
This moreover establishes a fundamental connection between multipartite entanglement and many-body correlations contained in response functions, with profound implications close to quantum phase transitions. There, the quantum Fisher information becomes universal, allowing us to identify strongly entangled phase transitions with a divergent multipartiteness of entanglement. We illustrate our framework using paradigmatic quantum Ising models, and point out potential signatures in optical-lattice experiments. 
\end{abstract}

\maketitle

\date{\today}


Entanglement is a central theoretical concept underlying the characterisation of quantum many-body states in condensed-matter and high-energy physics, as well as quantum information. For example, entanglement properties reveal exotic states of matter such as topological spin liquids \cite{Balents2010} or many-body localization \cite{Nandkishore2014,Smith2015}, the holographic entanglement entropy identifies confinement/deconfinement transitions in gauge theories \cite{Nishioka2007,Klebanov2008}, and entanglement is considered the central resource for quantum-enhanced metrology \cite{Escher2011,Pezze2014} as well as quantum computation \cite{Ladd2010,Cirac2012,Hauke2011d,Georgescu2014}. 
In experiments, entanglement becomes measurable via a tomographic determination of the many-particle quantum state \cite{Kim2010,Jurcevic2014,Lanting2014,Fukuhara2015}, 
and protocols have been developed \cite{Daley2012} and implemented in remarkable experiments \cite{Islam2015} to measure entanglement entropies in quench dynamics and quantum phase transitions. 
However, the resources required by these protocols scale exponentially with the system size, and these experimental efforts are thus limited a priori to few-particle systems. 

\begin{figure}
\centering
\includegraphics[width = 0.995\columnwidth]{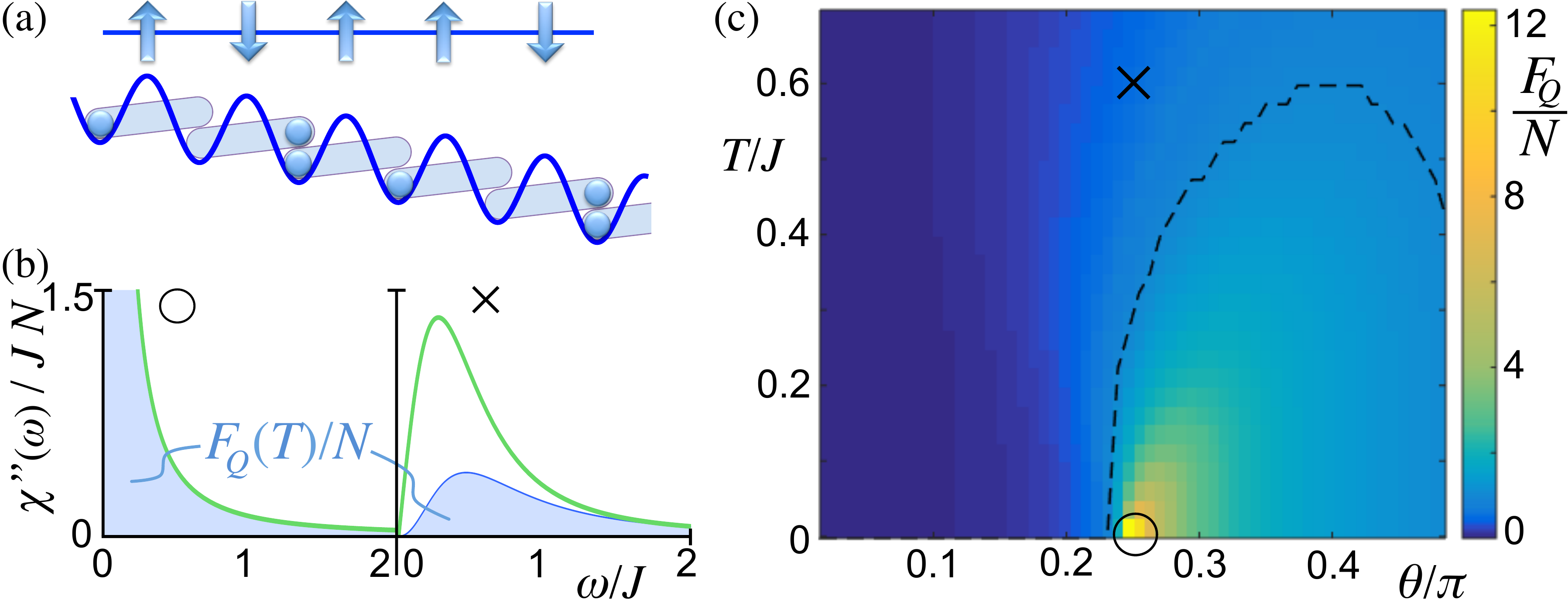}
\caption{(color online) Measurement prescription for the quantum Fisher information, a witness for multipartite entanglement. 
{\bf (a)} A quantum many-body system is prepared in a thermal state at temperature $T$. 
The example shows an Ising spin chain (top), Eq.~\eqref{eq:Ising}, as realisable in optical-lattice experiments (bottom) \cite{Simon2011}, 
but the general concept applies also to fermions and bosons, and in any spatial dimension.  
{\bf (b)} Standard tools, such as inelastic Bragg or neutron scattering, measure the imaginary part of the dynamic susceptibility $\chi''(\omega,T)$ as a function of frequency exchange $\omega$ (green line). 
The integral ${F_Q(T)} = \frac{4}{\pi} \int_{0}^{\infty} \ud \omega \tanh\left(\frac{ \omega}{2T}\right) \chi''(\omega,T)$ gives the quantum Fisher information (shaded areas). 
{\bf (c)} 
This procedure allows mapping out the quantum Fisher information as a function of temperature and transverse field $\theta$. 
Low-temperature states (circle) can host entanglement, but it is lost at larger temperatures (cross). 
The quantum Ising chain has a divergence of entanglement close to the quantum critical point, as well as a robust entangled region extending to finite temperatures (enclosed by the dashed line).   
}
\label{fig:concept}
\end{figure}

To address the problem of detecting and quantifying multipartite entanglement for large systems, we consider below the quantum Fisher information (QFI) as an entanglement witness \cite{Hyllus2012,Toth2012,Strobel2014}. Our key result is that---for a many-body system at thermal equilibrium at any temperature---the QFI can be determined directly  from a measurement of Kubo linear response functions, in particular the dynamic susceptibility (see Fig.~\ref{fig:concept}). We emphasise that this measurement prescription is independent of microscopic details of the system of interest and that the measurement of linear response is a standard tool in experiments. Importantly, only modest measurement resources are required that do not scale with system size. The presented prescription therefore makes multipartite entanglement observable for a large variety of experimental platforms, including quantum degenerate atomic gases as well as condensed-matter systems.

The discovered connection between QFI and response functions has profound implications for our understanding of quantum many-body systems. For example, the quantum fluctuations and quantum correlations contained in response functions provide a rigorous lower bound to multipartite entanglement in thermal ensembles. 
Even more, this intimate connection to response functions immediately implies universality of the QFI close to quantum phase transitions (QPTs), as has been observed theoretically in ground-state phase diagrams \cite{Ma2009,Liu2013QuantumFisher,Wang2014,Zheng2015} 
and is known for related metrics 
\cite{Gu2010}.  
Below, we discuss the universal scaling laws of the QFI at vanishing as well as non-zero temperature, and use them to identify a class of strongly-entangled QPTs with a divergent multipartiteness of entanglement and an associated divergent entanglement length scale \cite{Verstraete2004c}.  

This article is organised as follows. 
After outlining background on the QFI, we present the main result of this work, the relation of the QFI to Kubo response functions. 
Then, we discuss its universal scaling theory, and discuss why QPTs yield strong divergencies while thermal phase transitions do not present particular signatures. 
We illustrate these completely general concepts by paradigmatic examples. 
Finally, we discuss some practical experimental considerations, and provide our conclusions. 
\\ 

\emph{Background on the quantum Fisher information.---}
In recent years, the QFI has generated a lot of attention, because it provides a rigorous lower bound for genuinely multipartite entanglement  \cite{Hyllus2012,Toth2012}. 
Originally, it was introduced to quantify the maximal precision with which a parameter (a phase) $\vartheta$ can be estimated using a given quantum state $\rho$~\cite{Pezze2014,BraunsteinCaves1994}. 
For $M$ independent measurements, the quantum Fisher information $F_Q$ bounds the variance of $\vartheta$ by $(\Delta\vartheta)^2\geq 1/(M F_Q)$, the so-called quantum Cram\'er--Rao bound  \cite{BraunsteinCaves1994}. 

Importantly, the maximal precision achievable in quantum phase estimation can break classical limits if the employed system is in a strongly entangled $N$-particle state \cite{Pezze2014}.   
More precisely (see Supplementary Material \cite{supmatQFI}), if a state achieves a QFI density 
\be
	\label{eq:QFIDensityandEntanglement}
	 f_Q \equiv F_Q/N > m \,,
\ee
with $m$ a divisor of $N$, then it must contain $m+1$-partite entanglement \cite{Hyllus2012,Toth2012}. 

In quantum phase estimation, the QFI determines the sensitivity of the state $\rho$ towards a unitary transformation generated by the hermitian operator $\obs$ associated to $\vartheta$. That is, it quantifies the distinguishability of $\rho$  from $\rho'=\ue^{-i\delta\vartheta \obs }\rho\,\ue^{i\delta\vartheta \obs }$. 
For a pure quantum state $\rho = |\psi \rangle \langle \psi|$, such as the ground state of a given Hamiltonian, the QFI assumes the simple form of a connected correlation function, which can be easily computed or measured, 
\be
	F_Q = \Delta(\obs)^2 = \langle \psi | \obs \obs | \psi \rangle - \langle \psi | \obs |\psi \rangle^2\,. \label{eq:QFIpure}
\ee

Matters become much more complicated in a mixed state, such as a thermal ensemble $\rho=\sum_\lambda p_\lambda |{\lambda}\rangle\langle{\lambda}|$, where $\ket{\lambda}$ is the energy eigenbasis  with occupation probabilities $p_\lambda=\exp{(-E_\lambda/T)}/Z$, and $Z$ is the partition function. In such a case, the QFI takes the considerably more complex structure  
\be
	\label{eq:QFI}
	F_Q = 2\sum_{\lambda,\lambda^\prime} \frac{(p_\lambda-p_{\lambda^\prime})^2}{p_\lambda+p_{\lambda^\prime}} \left|\bra{\lambda}\obs\ket{\lambda^\prime}\right|^2   
\ee
(where the sum includes only terms with $p_\lambda+p_{\lambda^\prime}>0$).   
Recently, in a remarkable atomic-gas experiment~\cite{Strobel2014}, it has been demonstrated that a lower bound on the QFI can be measured by studying the behaviour of an observable's probability distribution under the unitary transformation $\ue^{i\delta\vartheta \obs }$. This could then be used to demonstrate the presence of bipartite entanglement. 
Here, we show how the QFI can be measured directly and efficiently in a generic quantum many-body system in a thermal state at any temperature.\\ 

\emph{Main result.---}
As the major result of this work, we rigorously relate the QFI to a Kubo response function, 
\be 
	\label{eq:FQImChiIntro}
	{F_Q(T)} = \frac{4}{\pi} \int_{0}^{\infty} \ud \omega \tanh\left(\frac{ \omega}{2T}\right) \chi''(\omega,T) \,, 
\ee
where $\chi''(\omega,T)=  \Im(\chi(\omega,T))$ is the imaginary, dissipative part of the dynamic susceptibility in the state $\rho$ with respect to $\obs$---the same thermal state and generator for which the QFI is evaluated 
\footnote{Interestingly, similar correlations, but in imaginary time, are being used to make the fidelity susceptibility calculable in quantum Monte Carlo computations. This computational technique also bounds the QFI (see Ref.~\cite{LeiWang2015} and references therein). Here, by relating the QFI to response functions in real time, we obtain a tool to measure it directly in laboratory experiments.}. 

\emph{Proof:} The proof of Eq.~\eqref{eq:FQImChiIntro} is straightforward. It requires only the minimal assumption of thermal equilibrium. Setting $\hbar=1=k_B$, the dynamic susceptibility is defined as   
\be
	\label{eq:susceptIntro}
	\chi(\omega,T) = i \int_{0}^\infty \ud t\, \ue^{-i \omega t}\,\mathrm{tr} \left(\rho \left[\obs(t),\obs\right] \right)\,,
\ee
where $\obs(t)=\ue^{i H t}\obs\ue^{-i H t}$. 
It is convenient to work in the Lehmann representation, i.e., the energy eigenbasis, where  
\be
	\chi''(\omega) = \sum_{\lambda,\lambda^\prime} (p_\lambda-p_{\lambda^\prime}) \left|\bra{\lambda}\obs\ket{\lambda^\prime}\right|^2   \pi \delta(\omega + E_{\lambda^\prime}-E_{\lambda}) \,.
\ee
Exploiting that for a thermal state $2 \int_0^\infty \ud\omega \, \tanh\left(\frac{\omega}{2T}\right) \delta(\omega + E_{\lambda^\prime}-E_{\lambda})=\tanh\left(\frac{E_{\lambda^\prime}-E_{\lambda}}{2T}\right)=\frac{p_\lambda-p_{\lambda^\prime}}{p_\lambda+p_{\lambda^\prime}}$, and correcting for prefactors, we directly obtain Eq.~\eqref{eq:FQImChiIntro}. 
This proof can be straightforwardly extended to the QFI matrix, and, via the fluctuation-dissipation theorem $\chi''(\omega) = \tanh(\omega/2T) S(\omega)$, to the dynamic structure factor $S(\omega)$. 
Equation~\eqref{eq:FQImChiIntro} also presents some direct corollaries, such as a sum rule which we discuss in the Supplementary Material \cite{supmatQFI}. 
$\Box$

The conceptual importance of the identification~\eqref{eq:FQImChiIntro} is huge. 

First of all, it makes a witness for multipartite entanglement, the QFI, a straightforwardly measurable quantity.  
Dynamic susceptibilities are routinely measured in many-body systems using well-established techniques such as Bragg spectroscopy 
\cite{Stoferle2004,Ernst2010} 
or neutron scattering \cite{Parker2013}. 

Second, the central Eq.~\eqref{eq:FQImChiIntro} has fundamental theoretical implications. 
For example, it establishes a direct relation between quantum correlations contained in $\chi''(\omega,T)$ to many-body entanglement. 
As Eq.~\eqref{eq:FQImChiIntro} shows, the entanglement contained in $\chi''(\omega,T)$ and extracted by the QFI is dominated by the thermally unaccesible high-frequency response. 

Third, the connection~\eqref{eq:FQImChiIntro} has especially profound consequences near continuous QPTs when choosing for $\obs$ a relevant operator in the renormalisation-group sense, such as the order parameter. Then, known universal scaling laws for $\chi''(\omega,T)$ translate directly into universal scaling for $F_Q$. 
Universal scaling of the QFI has already been theoretically observed in ground states of many-body systems \cite{Ma2009,Liu2013QuantumFisher,Wang2014,Zheng2015}. In the following, we extend the analysis to the experimentally relevant regime of non-zero temperatures.\\ 

\emph{Universal scaling of multipartite entanglement.---}
Consider a local generator $\obs=\sum_{l=1}^N \obs_l$, in a $d$-dimensional system with linear size $L$ and $N=L^d$ sites. 
As explained in the Supplementary Material \cite{supmatQFI}, $f_Q$ obeys the universal behaviour
\begin{equation}
	 f_Q(T,L^{-1},\tilde{h})=\lambda^{\Delta_Q}{\phi}_Q(T \lambda^z, L^{-1}\lambda,  \lambda^{1/\nu}\tilde{h} )\,.
	 \label{eq:sca_fisher_maintext}
\end{equation} 
Here, $z$ is the dynamical and $\nu$ the correlation-length critical exponent. $\lambda$ is the cutoff scale determined by the relevant perturbations $L^{-1}$, $T$, and the distance from the critical point $\tilde{h}$. 
The scaling dimension of $f_Q$ is $\Delta_Q=d-2\Delta_{\alpha}$, with $\Delta_\alpha$ the scaling dimension of $\obs_l$.  

Since $f_Q$ bounds the number of entangled particles via Eq.~\eqref{eq:QFIDensityandEntanglement}, its scaling behaviour allows us to identify a class of  strongly entangled QPTs, i.e., QPTs with a  divergent multipartiteness of entanglement. These are those transitions with $\Delta_Q>0$. 
The scaling behaviour additionally implies a universal length scale over which multipartite entanglement exists, $l_{\mathrm{ent}}\gtrsim f_Q^{1/d} \sim \lambda^{1-2\Delta_\alpha/d}$ (see Supplementary Material \cite{supmatQFI}).


\begin{figure}
\centering
\includegraphics[width = 0.7\columnwidth]{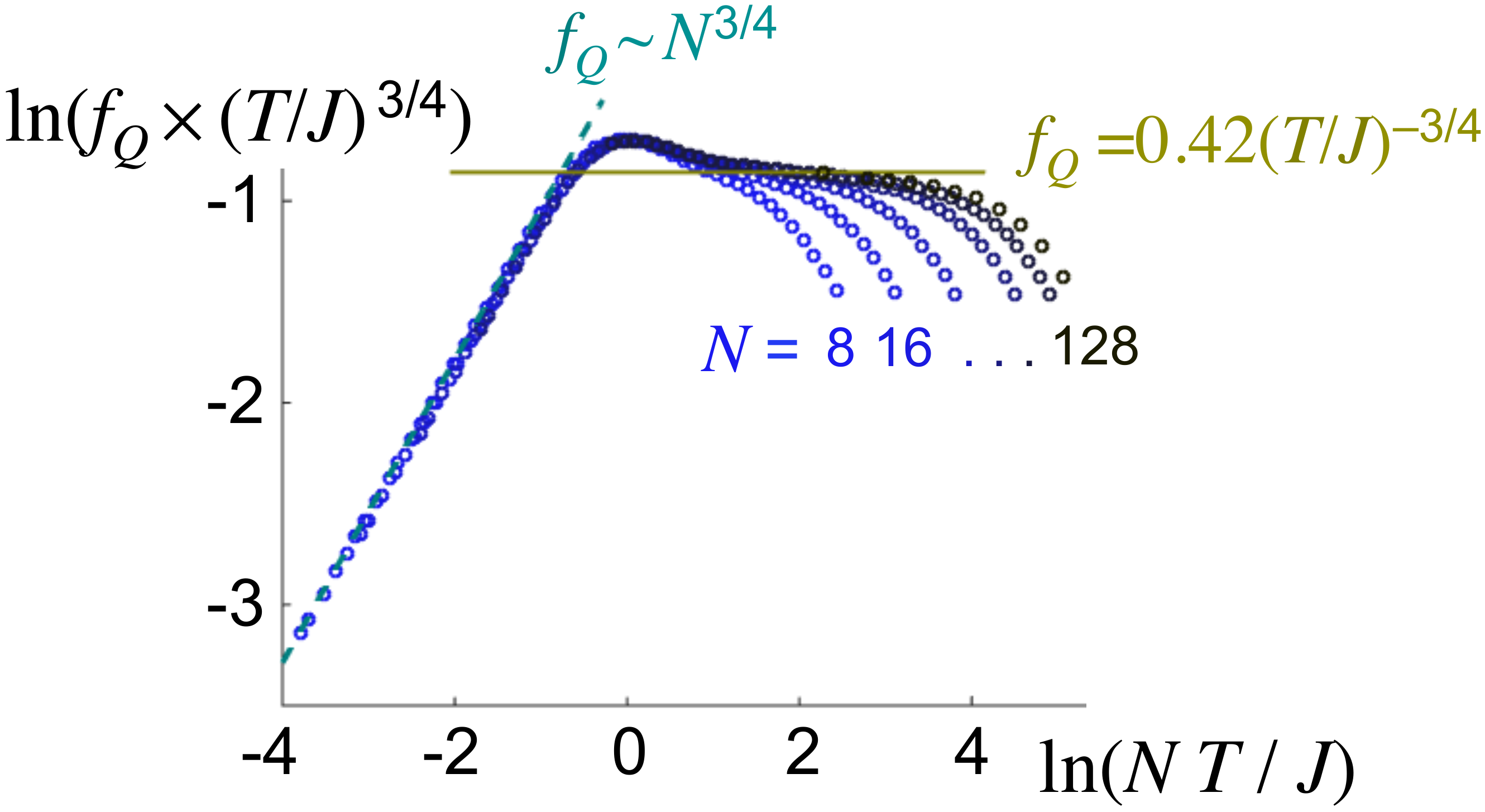}
\caption{(color online) Universal scaling of the quantum Fisher information density, $f_Q=F_Q/N$, calculated for the order parameter in the quantum Ising chain. 
Data at the critical point and for $N=8\dots 128$. 
At low temperatures, the universal scaling laws produce an excellent data collapse. 
In a range of intermediate temperatures, the scaling approaches analytic predictions for the thermal critical regime in the thermodynamic limit~\cite{Sachdev1999}  (solid line). 
The strong divergence at small temperatures of $f_Q\sim N^{3/4}$ (dashed line) implies a diverging multipartiteness of entanglement. }
\label{fig:FQIsing}
\end{figure}


The framework described up to now is completely general. To illustrate its viability for understanding entanglement in quantum many-body systems, we now turn to specific examples. 
We focus on a paradigmatic class of model systems presenting a QPT, namely Ising models in a transverse field, which are realisable in quantum-optical \cite{Simon2011,Strobel2014} 
as well as solid-state systems \cite{Sachdev1999,Haelg2015}, 
\be
	\label{eq:Ising}
	\frac{H}{J} = - \cos\theta \sum_{l,j=1}^{N} \mathcal{J}_{lj} \sigma_l^x \sigma_{j}^x  + \sin\theta \sum_{l=1}^N \sigma_l^z\,. 
\ee
Here, $\sigma_l^\alpha$, $\alpha=x,y,z$, are the Pauli matrices on lattice sites $l=1,\dots,N$. 
Depending on the Ising interactions $\mathcal{J}_{lj}$, this Hamiltonian exhibits a quantum critical point at some critical field strength $\theta_c$. 
The order parameter for the transition is $\sum_l\braket{\sigma_l^x}/N$. Its strong critical fluctuations make the generator $\obs=\sum_l\sigma_l^x /2$ an ideal candidate for testing the scaling behaviour of the QFI. 

We first focus on the simplest case, the one-dimensional nearest-neighbour Ising chain, $\mathcal{J}_{lj}=\delta_{j,l+1}$, where the scaling exponents are known analytically, $z=1$ and $\Delta_\alpha=1/8$ \cite{Sachdev1999}. 
This gives $\Delta_Q=3/4$, i.e., the nearest-neighbour quantum Ising chain lies in the class of strongly entangled phase transitions with divergent multipartite entanglement. Indeed, $f_Q$ for the order parameter displays a strong peak around the critical point $\theta_c=\pi/4$, see Fig.~\ref{fig:concept}(c). 
The entanglement radiates out from the peak, generating a broad entangled region also at non-zero $T$ \cite{Toth2005,Wu2005}. 

To illustrate the Ansatz~\eqref{eq:sca_fisher_maintext}, we consider the scaling with system size $L$ and temperature $T$, at fixed $\tilde{h}=0$. 
For small $L$, the dominant cutoff scale is $\lambda\sim L=N$, implying 
\be
	\label{eq:scalingwithsize_Isingchain}
	f_Q \sim N^{3/4}\,.
\ee
The data in Fig.~\ref{fig:FQIsing} reproduces perfectly this strong algebraic growth, which is remarkably close to the theoretical maximum of $f_Q=N$. The associated multipartite entanglement length scale is thus highly divergent, $l_{\mathrm{ent}}\sim N^{3/4}$. 

With increasing temperature, the cutoff scale crosses over to $\lambda\sim T^{-1/z}$, and the scaling becomes $f_Q \sim T^{-3/4}$. 
This prediction can be refined by exact analytical results for the dynamic susceptibility at criticality \cite{Sachdev1999}, 
$\lim_{N\to\infty}{\chi(\omega,T)}/{N} = g(\omega/T) {J}^{3/4}/{T}^{7/4} $, with a function $g$ that is known exactly.  
Performing the $\omega$-integral in Eq.~\eqref{eq:FQImChiIntro}, one immediately obtains the QFI, 
\be 
	\label{eq:FQIsingInfty}
	f_Q(T) =  C\left({J}/{T}\right)^{3/4}\,, 
\ee
with $C\approx 0.42$. In the temperature regime of validity, $N T/J\gg 1$ and $T/J\ll 1$, the exact data for finite chains is consistent with this scaling prediction valid in the thermodynamic limit (see Fig.~\ref{fig:FQIsing}). 
For $T\gg J$ the system crosses over into a generic high-temperature asymptotic behaviour $f_Q\sim T^{-2}$.\\


\emph{Absence of signature at thermal phase transitions.---}
Remarkably, such scaling behaviour is only observed at quantum, but not thermal phase transitions, because Eq.~\eqref{eq:FQImChiIntro} considers only quantum fluctuations. 
A simple example to demonstrate the insensitivity towards thermal phase transitions is provided by the fully-connected transverse-field Ising model, $\mathcal{J}_{lj}=1/N$, $\forall\, {l,j}$, similar to the model describing the experiments of Ref.~\cite{Strobel2014}. 
In contrast to its nearest-neighbour counterpart, this model exhibits, additionally to the QPT at $\theta_c=\pi/4$, also a thermal phase transition \cite{Das2006}.

\begin{figure}
\centering
\includegraphics[width = 0.995\columnwidth]{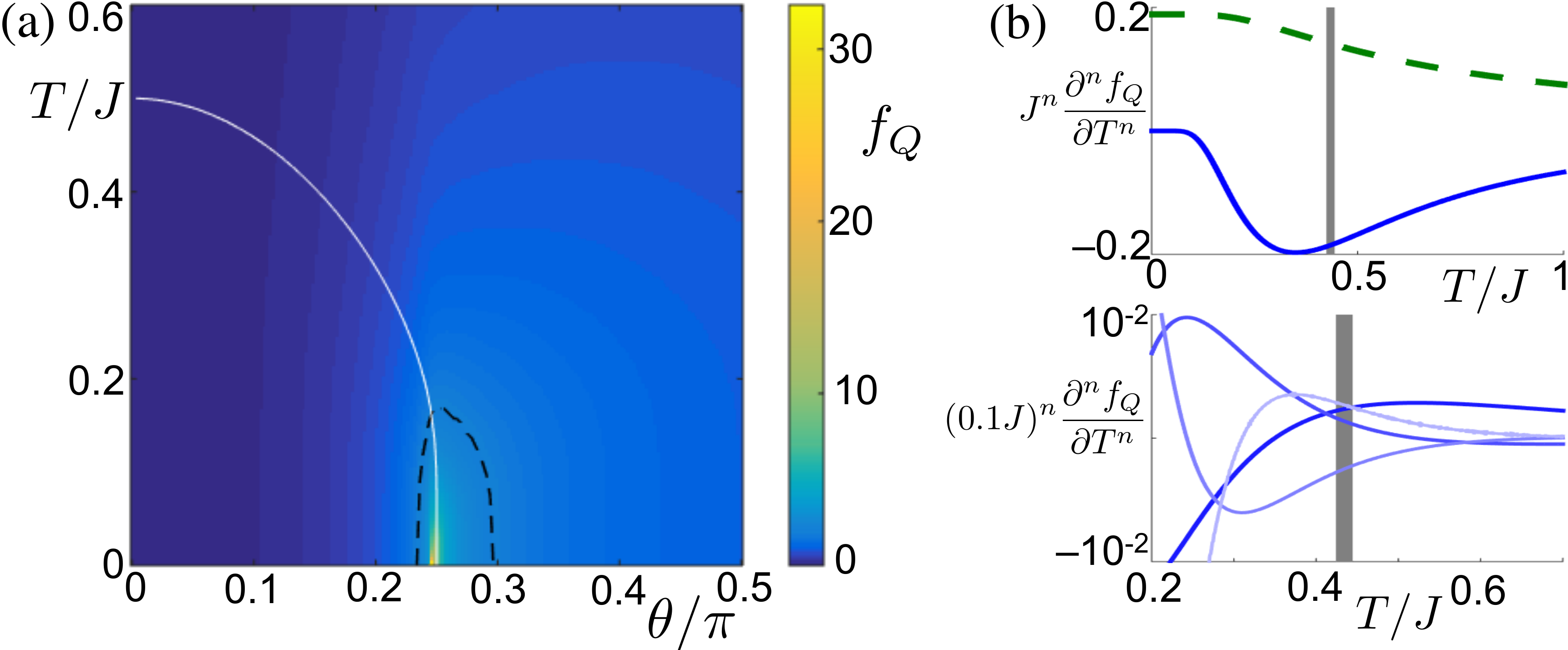}
\caption{(color online) Absence of features of the quantum Fisher information at thermal phase transitions, exemplified by the infinitely-connected Ising model. 
{\bf{(a)}} 
Finite entanglement is witnessed at non-zero temperatures (enclosed by the dashed line), and a divergence appears near the quantum critical point at $\theta=\pi/4$ and $T=0$. 
But, crucially, $f_Q$ shows no features at the thermal phase transition (solid line). Data for $N=1000$. 
{\bf{(b)}} Derivatives of $f_Q$ remain also smooth at the thermal transition (vertical line). 
Top: $f_Q$ (dashed) and $\partial f_Q/\partial (T/J)$ (solid line). 
Bottom, from dark to light: second to fifth derivative. 
Data for $N=1600$ and $\theta=\pi/8$. 
}   
\label{fig:FQ_LRIsing}
\end{figure}

Figure~\ref{fig:FQ_LRIsing}(a) shows $f_Q$ for the order parameter in the temperature--transverse-field plane. We delegate its scaling analysis to the Supplementary Material \cite{supmatQFI}. 
More important at this point, while $f_Q$ shows a divergence at the QPT, no particular feature can be discerned at the thermal phase transition. Neither do such features appear in derivatives of $f_Q$ [Figure~\ref{fig:FQ_LRIsing}(b)]. 

However, it is known that the \emph{static} isothermal susceptibility $\chi^T$ of the order parameter diverges \cite{bookCardy}. 
The reason why the QFI instead remains featureless becomes clear when decomposing $\chi^T=\chi_{\mathrm{el}}+\chi_{\mathrm{vV}}$ into its two fundamental parts  \cite{JensenBookChapter3}, the elastic (or Curie) contribution $\chi_{\mathrm{el}}$, and the quantum-mechanical van-Vleck correction $\chi_{\mathrm{vV}}$, which is continuously connected to the Kubo susceptibility, $\chi_{\mathrm{vV}}=\lim_{\omega\to 0}\chi(\omega,T)$. 
It is $\chi_{\mathrm{el}}$ that diverges at a thermal phase transition. 
As is shown in the Supplementary Material \cite{supmatQFI}, $\chi_{\mathrm{el}}$ can be related to the Fisher information in a classical scenario \cite{ProkopenkoObst2011} that has no relation to entanglement. 
The second term $\chi_{\mathrm{vV}}$, on the other hand, remains smooth at thermal transitions. 
The QFI, thus, considers only the contribution to the susceptibility that is due to quantum fluctuations and remains insensitive to thermal phase transitions.\\  

\emph{Experimental considerations.---}
Let us finally address some practical aspects that will be important for experiments.  
The measurement prescription~\eqref{eq:FQImChiIntro} is very flexible, since its proof did not make any assumptions on microscopic details of the system under study (other than thermal equilibrium). 
As a consequence, it applies in any spatial dimension, for any hermitian generator $\obs$, and it can be equally used for systems of spins, bosons, or fermions. 
For example, the quantum Fisher information may also be probed in ongoing atomic-gas experiments that aim at studying low-temperature phases of the Fermi--Hubbard model~\cite{Greif2013,Hart2015}. 
A bound for multipartite spatial entanglement as in Eq.~\eqref{eq:QFIDensityandEntanglement} is known to exist in all cases where the generator is a sum of local operators, $\obs=\sum_l \obs_l$, when $\obs_l$ has a bounded spectrum \cite{Pezze2014} (see Supplementary Material \cite{supmatQFI}). 

Importantly, the complexity of measuring $\chi''(\omega,T)$ does not scale with system size, thus allowing for an efficient evaluation of the QFI via Eq.~\eqref{eq:FQImChiIntro}. 
Moreover, the scaling analysis overcomes a usual practical difficulty for studying the QFI, the optimal choice for the linear generator $\obs=\sum_l \obs_l$. 
In the vicinity of a QPT, the choice becomes clear: one may select any suitable, relevant operator $\obs_l$, preferably the one with the largest scaling exponent $\Delta_Q$. 
A good choice will often be the order parameter, as in the examples above. 
The Supplementary Material~\cite{supmatQFI} contains an example for the Mott-insulator--superfluid transition, where a generator different from the order parameter but accessible by Bragg spectroscopy, allows one to extract universal behaviour.

Further, the prescription is robust against intrinsic sources of imperfections. 
For example, an uncertainty in determining the system's temperature can be mitigated by choosing the lowest temperature estimate. Due to the monotonicity of the $\tanh(\omega/2T)$ as a function of frequency, this produces a lower bound for the QFI. 
Another natural error source is a finite spectral resolution. Assuming a symmetric spectral broadening, the concavity of the $\tanh$ assures that the integration again returns a lower bound to the true QFI, with less severe results at smaller temperatures. 
Also, the integral in Eq.~\eqref{eq:FQImChiIntro} will be limited to some frequency range (the upper integral limit  represents, as usual, a scale much larger than energies accessible by the considered degrees of freedom). This limitation will again deliver a lower estimate. 
Importantly, non of these errors will produce a false positive indicator for entanglement. 
Even more, close to a QPT, the existence of the universal scaling laws assures an inherent robustness of the QFI. 
Therefore, in contrast to some engineered highly-entangled states, such as the GHZ state, the entanglement witnessed by the QFI close to QPTs is unaffected by weak symmetry-preserving perturbations.\\


\emph{Conclusion.---}
To conclude, we have developed a simple prescription for measuring the quantum Fisher information, which is independent of spatial dimensionality or other microscopic details, and relies only on well-developed tools for measuring dynamic susceptibilities. 
Many-particle entanglement thus becomes observable in various experimental platforms, including quantum degenerate atomic gases as well as solid-state samples. 

The discovered prescription adds an important tool to the ongoing search for measurable entanglement witnesses \cite{Amico2008,Guehne2009}, some of which have already allowed experiments to demonstrate the presence of bipartite entanglement~\cite{Ghosh2003,Brukner2006,Vertesi2006,Cramer2013,Strobel2014}. 
The prescription is also complementary to efforts for measuring the entanglement entropy \cite{Islam2015}, which can reveal more complete information, at the expense, however, of an exponential increase in measurement resources. Moreover, generic scaling behaviour of the block entanglement entropy is rigorously demonstrated only near critical points that are described by a conformal field theory \cite{Amico2008}. 
In contrast, the QFI displays scaling behaviour in the vicinity of arbitrary continuous quantum phase transitions. 
This scaling behaviour allowed us to classify quantum phase transitions with a divergent multipartiteness of entanglement.  

In order to achieve an advantage over classical systems, identifying and quantifying large-scale entanglement in many-particle systems will be an indispensable ingredient for quantum simulation \cite{Cirac2012,Hauke2011d,Georgescu2014} and quantum metrology \cite{Escher2011,Pezze2014}.  
It will therefore be worthwhile for future studies to generalise the concept developed here to other kinds of phase transitions, e.g., topological, or to states beyond thermal ensembles, such as generalised Gibbs ensembles.\\


\emph{Acknowledgements.---}
We thank J.~Ignacio Cirac, Rosario Fazio, Luca Pezz\'e, and Augusto Smerzi for stimulating discussions. 
We acknowledge support from 
the EU IP SIQS, SFB FoQuS (FWF Project No.~F4016-N23), the ERC synergy grant UQUAM, the Deutsche Akademie der Naturforscher Leopoldina (grant No.~LPDS 2013-07 and LPDR 2015-01), Spanish Government Grant FOQUS, ERC AdG OSYRIS, EU STREPEQuaM, and EU FET Proactive QUIC. 
The evaluation of the Pfaffians required for calculating dynamic  susceptibilities in the one-dimensional models uses the algorithm provided in Ref.~\cite{Wimmer2012}.
\\


\newpage

\renewcommand{\theequation}{S\arabic{equation}}
\setcounter{equation}{0}
\renewcommand{\thefigure}{S\arabic{figure}}
\setcounter{figure}{0}
\renewcommand{\thesection}{S\arabic{section}}
\setcounter{section}{0}

{\begin{center}{\LARGE{\bf Supplementary material}}\end{center}}

In this supplementary material, 
we 
(i) provide additional technical details, 
(ii) explain in detail the scaling behaviour for the quantum Fisher information and present additional numerical results,    
(iii) discuss the relationship Fisher-information/susceptibility in a classical, thermal setting, and
(iv) present a sum rule that may be used to bound the quantum Fisher information.

\section{Relation of QFI to multipartite entanglement \label{sec:Entanglement}}

The QFI has a deep connection with multipartite entanglement~\cite{Hyllus2012,Toth2012}. 
Consider a system of $N$ spins with length $S=\frac 1 2$, and a linear observable $\obs = \frac{1}{2}\sum_l {\mathbf{n}_l}\cdot \boldsymbol\sigma_l$, where $\boldsymbol\sigma_l=\left(\sigma_l^x,\sigma_l^y,\sigma_l^z\right)$ is a vector containing the Pauli matrices $\sigma_l^\alpha$ associated to spin $l$, and $\mathbf{n}_l=(n_l^x,n_l^y,n_l^z)$ is a unit vector on the Bloch sphere.  
Then the system hosts at least $m+1$-partite entanglement if the QFI associated to $\obs$ fulfils \cite{Hyllus2012}
\be
	\label{eq:QFIandEntanglement}
	F_Q > \left\lfloor \frac{N}{m} \right\rfloor m^2 + \left(N - \left\lfloor \frac{N}{m} \right\rfloor m\right)^2 \,,
\ee
where $\left\lfloor X \right\rfloor$ is the largest integer smaller than or equal to $X$. 
For $m$ a divisor of $N$, the condition \eqref{eq:QFIandEntanglement} attains the simple form \eqref{eq:QFIDensityandEntanglement} when expressed through the quantum Fisher information density $f_Q\equiv F_Q/N$. 
Note that typically response functions that measure linear operators contain a factor of $1/N$ relative to our definition, Eq.~\eqref{eq:susceptIntro}, so that they will observe directly $f_Q$ rather than $F_Q$. 

The proof of Ref.~\cite{Hyllus2012} for bounding the multipartite entanglement with Eq.~(\ref{eq:QFIandEntanglement}) can be directly translated to degrees of freedom other than spins $1/2$, as long as $\obs$ represents a sum of local operators with bounded spectrum \cite{Pezze2014}. If $h_{\mathrm{max}}$ and $h_{\mathrm{min}}$ denote the largest, respectively smallest, eigenvalue of $\obs$,   then the right hand side of the condition \eqref{eq:QFIandEntanglement} acquires the prefactor $(h_{\mathrm{max}}-h_{\mathrm{min}})^2$. 
Therefore, the QFI can also witness spatial entanglement in systems other than spins-$1/2$, such as larger spins or fermions.  
Note that the relation of the QFI to response functions, Eq.~\eqref{eq:FQImChiIntro}, is independent of any such microscopic details of the underlying quantum many-body system or the hermitian operator $\obs$, which may even be non-local or unbounded.\\

\section{Solvability of the considered models \label{sec:solvability}}

All models used in this article for illustrating the main concepts are exactly solvable. 
The one-dimensional Ising chain in a transverse field [Eq.~\eqref{eq:Ising} with $\mathcal{J}_{lj}=\delta_{j,l+1}$] can be mapped to a free-fermion problem \cite{Sachdev1999}. 
Dynamical susceptibilities can then be calculated via Wick's decomposition of expectation values \cite{Derzhko1997}, and the Pfaffians appearing in the resulting expressions can be evaluated efficiently using the algorithm described in Ref.~\cite{Wimmer2012}.

The infinite-range Ising Hamiltonian [Eq.~\eqref{eq:Ising} with $\mathcal{J}_{lj}=1/N$] commutes with both $\mathbf{S}^2 = (S^x)^2 + (S^y)^2 + (S^z)^2$ and $S^z$ where $S^\alpha=\sum_{l} \sigma_l^\alpha/2$, $\alpha=x,y,z$. As a consequence, the Hamiltonian decomposes into disconnected blocks when represented in the common eigenbasis of $\mathbf{S}^2$ and $S^z$. Each block grows linearly with particle number $N$ and can be diagonalized efficiently. Here, we consider the largest of these blocks, with dimension $N+1$. 
The infinite-range Ising model has a thermal phase transition for $\theta<\pi/4$, with critical temperature \cite{Das2006}   
\be
	\frac{T_c}{J}=\frac{\sin(\theta)}{\log[  (1+\tan(\theta)) / (1-\tan(\theta)) ]}\,.
\ee
In all models, we set $\hbar=k_B=a=1$, where $a$ is the lattice spacing.



\section{Scaling and universality \label{sec:Scaling}}

In the main text, we have demonstrated a close relationship of the QFI to thermodynamic response functions. 
These response functions underlie universal scaling behaviour close to quantum critical points, and consequently directly imprint a universal scaling behaviour onto the QFI. 
Aspects of critical scaling of the QFI and related quantum metrics have also been discussed, e.g., in 
Refs.~\cite{Gu2010,Ma2009,Liu2013QuantumFisher,Wang2014,Zheng2015}. 
The purpose of the present supplementary section is to present a general, unified scaling theory for the QFI. 

\subsection{Basic principles}

Before discussing the scaling of the QFI in detail, let us briefly recall the basic principles of the scaling Ansatz \cite{bookCardy}.  
Here, we focus on lattice models with lattice spacing $a$ that have a continuous phase transition at zero temperature, \emph{viz} a quantum phase transition (QPT). 
Generalisation to continuum models is straightforward.
The dimensionality of the system is $d$, its linear size $L$, and the total particle number $N=L^d$ (where we set $a=1$). 
To simplify the discussion, we focus on the simplest scenario, QPTs obtained by tuning a single relevant field, say the magnetic field $h$, to the critical point $h_c$. In this scenario, any small variation of $\tilde{h} = |h-h_c|$ drives the system in the thermodynamic limit to its stable fixed point with zero correlation length. Nevertheless, in the scaling regime, i.e., sufficiently close to the phase transition, all low-energy physics  becomes universal. As a remarkable consequence of this universality, very diverse microscopic models develop the same collective emerging behaviour at low energy, i.e., they display the same power-law decay of correlations, which is described by a small set of universal exponents $\set{\Delta_{\alpha}}$.

In finite systems, or systems at finite temperature, the power-law of the correlations is cut off by system size or the thermal correlation length. 
In such cases, one can extract the universal properties of the QPT by performing an appropriate analysis of scale transformations, i.e., transformations that modify the lattice spacing $a\to \lambda a$, where $\lambda>1$ (this scale transformation can be envisioned as grouping subsets of $\lambda$ spins together). 
Close enough to the critical point one can expand all local operators in the basis of scaling operators, i.e., those that transform under rescaling  with a simple power of $\lambda$, $\obs^{\alpha}\to \lambda^{-\Delta_{\alpha}}{\obs}^{\alpha}$, with scaling exponent $\Delta_{\alpha} \geq 0$. 
The  two-point correlation function of these operators decays as 
\be
\label{eq:scalingOfO}
\langle{{\obs}^{\alpha}(0){\obs}^{\alpha}(r)}\rangle-\langle{{\obs}^{\alpha}(0)}\rangle\langle{{\obs}^{\alpha}(r)}\rangle\propto r^{-2\Delta_{\alpha}}\,.
\ee
In particular, the fixed-point Hamiltonian is a scaling operator with scaling exponent $\Delta_{H}$.

Scaling operators that transform with $\Delta_{\alpha} > \Delta_{H}$  are called \emph{irrelevant} operators, since, when added as a weak perturbation, coarse graining decreases their importance relative to the original Hamiltonian. Those operators with  $\Delta_{\alpha} <  \Delta_{H}$, on the other hand,  are called  \emph{relevant}, since they become larger and larger under coarse-graining transformations, eventually driving the system to a different fixed point. 
In the \emph{scaling regime},  the regime where relevant perturbations are still small compared to the original Hamiltonian, one can extract the universal low-energy physics by studying the response of the system to scaling transformations. 

Both the inverse system size $L^{-1}$ and the finite temperature $T$ are  relevant perturbations. They increase under a rescaling as $L^{-1} \to \lambda L^{-1}$ and $T \to \lambda^z T$, where $z$ is the dynamical critical exponent. The scaling regime for a finite and cold system is thus defined by $L^{-1} \ll 1$ and $T \ll 1$.
We will now discuss the scaling behaviour of the QFI with respect to these perturbations, as well as to a deviation $\tilde{h}$ from criticality.

\subsection{Scaling of the quantum Fisher information}

To construct a scaling theory for the QFI close to a quantum critical point, we start from Eq.~\eqref{eq:QFIpure}, which defines the QFI at zero temperature, and consider a local generator $\obs=\sum_l {\obs}^\alpha_l$. 
By inserting Eq.~\eqref{eq:scalingOfO} into Eq.~\eqref{eq:QFIpure}, one finds that $f_Q$ transforms under rescaling by a factor $\lambda$ as 
\begin{equation}
 f_Q(T',L'^{-1},\tilde{h}')=\lambda^{d-2\Delta_{\alpha}}{\phi}_Q(T \lambda^z, L^{-1}\lambda,  \lambda^{1/\nu}\tilde{h} )+C\,.
 \label{eq:sca_fisher}
\end{equation}
Here, $C$ is a constant that is related to the non-universal short-distance correlations, and $\nu$ is the critical exponent associated to the correlation length. 
According to Eq.~\eqref{eq:sca_fisher}, the QFI has the scaling exponent $\Delta_Q = d-2\Delta_{\alpha}$. 
The scaling function ${\phi}_Q(T \lambda^z, L^{-1}\lambda,  \lambda^{1/\nu}\tilde{h} )$ encodes the large-distance part of the correlation functions, which dominates the scaling if the condition $\Delta_{Q}>0$ is fulfilled. In the following discussion, we assume that this is the case, allowing us to neglect the non-universal constant $C$, leading to Eq.~\eqref{eq:sca_fisher_maintext}. 
Below we will encounter, in the guise of the hard-core boson chain, an example where $\Delta_Q\leq 0$. In such cases, derivatives of $f_Q$ still allow one to extract the purely universal contribution. 
Note that when $\Delta_{Q}=0$ there can be logarithmic corrections to Eq.~\eqref{eq:sca_fisher}.  

We can use the scaling form Eq.~\eqref{eq:sca_fisher} as long as all relevant perturbations are small, that is as long as we stay in the scaling regime. 
The most relevant perturbation will constitute the most significant breaking of scale invariance, a fact that enables us to extract valuable information about the involved critical exponents. 
For example, when both $L^{-1}$ and $T$ are sufficiently small, the deviation $\tilde{h}$ from the critical point represents the most relevant perturbation: Scale invariance will be broken at the scale $\lambda\sim\tilde{h}^{-\nu}$. This implies that we can trade $\lambda$ by $\tilde{h}^{-\nu}$, and varying $\tilde{h}$ in this regime has the same effect as a rescaling transformation. In this regime, we expect $f_Q$ therefore to behave as 
\be
f_Q(T,L^{-1},\tilde{h})= \tilde{h}^{-\nu \Delta_{Q}}{\phi}_{Q1}\left(T \tilde{h}^{-\nu z}, L^{-1}\tilde{h}^{-\nu}\right)\,. 
\label{eq:fQscalingwithh}
\ee
For sufficiently relevant operators, for which $\Delta_Q >0$, $f_Q$ thus diverges  as $ \tilde{h}^{-\nu \Delta_Q}$ when approaching the critical point.
In the above expression, ${\phi}_{Q1}(T \tilde{h}^{-\nu z}, L^{-1}\tilde{h}^{-\nu})$ is a scaling function, embodying the fact that the normalised quantum Fisher information density, $f_Q/\tilde{h}^{-\nu \Delta_{Q}}$, is a well defined function of the scaling variables $ T \tilde{h}^{-\nu z}$ and $ L^{-1}\tilde{h}^{-\nu}$ rather than of $L$, $T$, and $\tilde{h}$  separately. 

Upon decreasing $\tilde{h}$, the system will reach a crossover with the next relevant perturbation. 
If $L^{-1} < T^{1/z}$, the crossover will occur around $T^{1/z} \sim \tilde{h}^{\nu}$ and carry over into a thermally dominated regime, for which the scaling of the QFI is dictated by 
\be
f_Q(T,L^{-1}, \tilde{h})= T^{-\Delta_Q/z}{\phi}_{Q2}\left(L^{-1} T^{-1/z},\tilde{h}T^{-1/(z\nu)}\right)\,.
\label{eq:fQscalingwithT}
\ee 
This scaling behaviour characterises the robustness of critical entanglement against non-zero temperatures (see, e.g., Ref.~\cite{Escher2011} for a discussion of the robustness in presence of noise).

In the opposite case $L^{-1} > T^{1/z}$, upon reducing $\tilde{h}$ one enters a regime dominated by finite-size effects. Here, $f_Q$ behaves as 
\be
\label{eq:fQscalingwithsize}
f_Q(T,L^{-1}, \tilde{h})= L^{\Delta_Q}{\phi}_{Q3}\left(T L^{ z},L^{1/\nu}\tilde{h}\right)\,.
\ee 
In this regime, $f_Q$ saturates to a finite, $L$ dependent value, which diverges with system size as $L^{\Delta_Q}$. The crossover is located around $L^{-1} \sim \tilde{h}^{\nu}$ and should thus approach the critical value $\tilde h=0$ as $L^{-1/\nu}$. 
Another interesting scenario appears when $T$ and $1/L$ are comparable, giving rise to further complex crossover phenomena. This can be seen in Fig.~\ref{fig:FQIsing} of the main text, where we study the universal scaling behaviour of the QFI in the paradigmatic model of the quantum Ising chain. 

This scaling analysis immediately permits us to identify strongly entangled quantum critical points. These are those critical points where $\Delta_Q>0$, i.e., where at low temperatures and close to criticality, $f_Q$ diverges with system size. 
Additionally, the scaling also allows us to extract a length scale of the entanglement present in the system \cite{Verstraete2004c}. 
Combining Eq.~\eqref{eq:sca_fisher} and main-text inequality~\eqref{eq:QFIDensityandEntanglement}, one sees that the number of entangled particles scales as $N_Q\geq f_Q \sim \lambda^{d-2\Delta_{\alpha}}$, which defines a multipartite entanglement length scale 
\be
	l_{\mathrm{ent}}\sim N_Q^{1/d}\gtrsim \lambda^{1-2\Delta_{\alpha}/d}\,.
\ee
Here, we assumed an isotropic distribution of entanglement; for anisotropic situations, the longest distance along which particles are witnessed to be entangled is actually larger. 
This entanglement length is related to the correlation length $\xi$ of the generator $\obs$. Since at criticality the correlation length scales as $\xi\sim \lambda$, one has $l_{\mathrm{ent}}\sim \xi^{1-2\Delta_\alpha/d}$. 
For $2\Delta_{\alpha}/d<1$, the length scale over which particles are entangled with each other thus diverges with system size.

\subsection{Universal scaling of in the Mott-insulator--superfluid transition}

The scaling behaviour of $f_Q$, Eq.~\eqref{eq:sca_fisher}, appears in all relevant operators, not only the order parameter. 
As an illustration for this fact, we choose a system that is relevant for experiments on cold gases, a system of ultracold bosonic atoms confined in a one-dimensional optical lattice. 
In the limit of strong on-site interactions, this system can be modelled by the Hamiltonian~\cite{Paredes2004} 
\be
	\label{eq:Hhardcorebosons}
	H_{\mathrm{hcb}} = - J \sum_{l=1}^{N-1} \left(\hb_l^\dagger \hb_{l+1} + \mathrm{h.c.} \right) - 2\mu \sum_{l=1}^N \hb_l^\dagger\hb_l \,,
\ee 
where the bosonic creation (annihilation) operators $\hb_l^\dagger$ ($\hb_l$) obey the hard-core constraint $\hb_l^\dagger \hb_l^\dagger=0$. 
This model has a QPT at $\mu_c/J=1$ from a critical superfluid phase to a Mott-insulator at uniform filling, with critical exponents $z=2$ and $\nu=1/2$ \cite{Sachdev1999}. 

The transition has several peculiarities. 
First, the exact ground state is fully separable at the QPT \cite{Giampaolo2008}. 
Second, one of its order parameters, the particle density $\hat{n}= \sum_l \hb_l^\dagger \hb_l/N$, commutes with the Hamiltonian. 
We can, however, use a modified observable that is still a relevant operator, the staggered density $\obs= \sum_l (-1)^l \hb_l^\dagger \hb_l$. 
In optical-lattice experiments, the dynamic susceptibility of such density fluctuations may be probed by Bragg spectroscopy \cite{Stoferle2004,Ernst2010}. 

\begin{figure}
\centering
\includegraphics[width = 0.995\columnwidth]{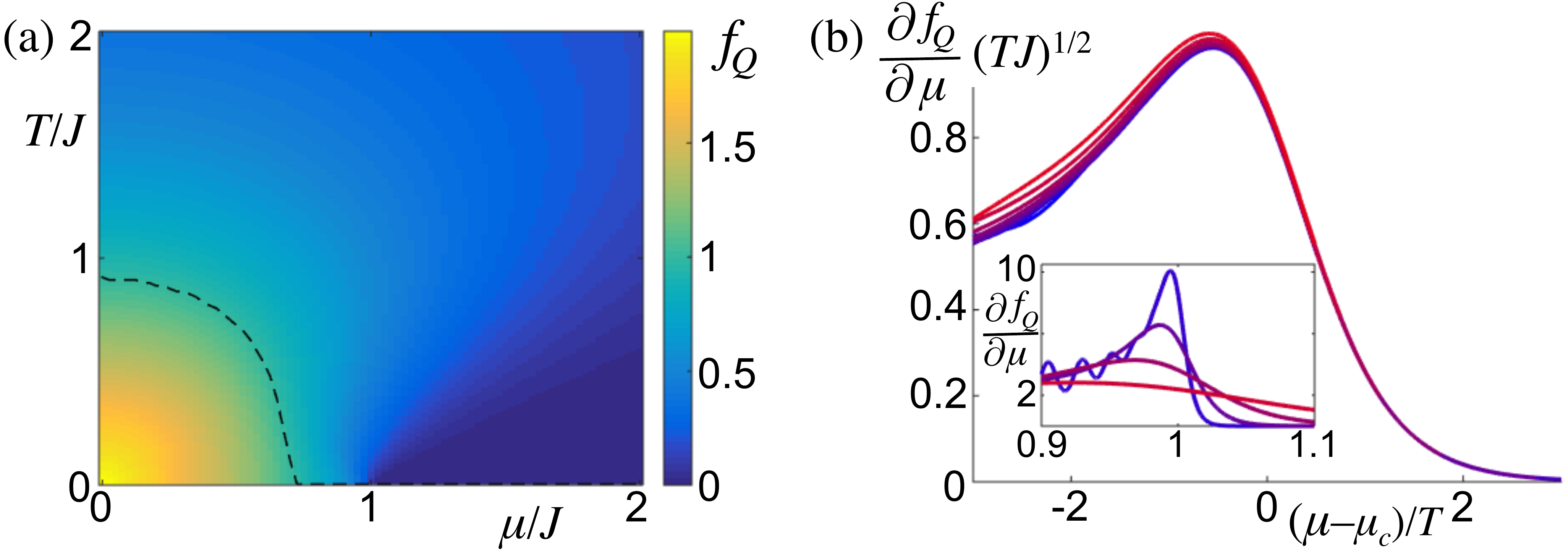}
\caption{
(color online) Quantum Fisher information density, $f_Q$, for the staggered density of hard-core bosons in a one-dimensional optical lattice. 
{\bf{(a)}} 
In part of the superfluid phase, $f_Q$ witnesses two-site entanglement. Data for $N=1000$. 
{\bf{(b)}} 
Inset: $\partial f_Q/{\partial \mu}$ shows a peak that, with decreasing $T$, diverges and moves towards the critical point. 
Main panel: The predicted scaling exponents produce a perfect data collapse for $\partial f_Q/{\partial \mu}$. 
Data for $N=100$ and, from blue to red, 8 different temperatures distributed on a logarithmic scale between $T=0.01-0.2 J$ (inset: 4 temperatures in $0.01-0.13 J$).  
}
\label{fig:FQXY}
\end{figure}

The Hamiltonian Eq.~\eqref{eq:Hhardcorebosons} is exactly solvable with the same methods as for the Ising chain in a transverse field \cite{Sachdev1999,Derzhko2000}. 
Moreover, under periodic boundary conditions, the dynamic structure factor $S(q,\omega)$ associated to density fluctuations $\obs=\sum_l \ue^{i q r_l} \hb_l^\dagger \hb_l$ at wave vector $q$ can be evaluated analytically \cite{Roux2013}. 
It is gapless at ${q}=0$ and acquires a maximum at ${q}=\pi$, where
\be
	S(\pi,\omega) = 2\pi \sum_k \delta(\omega -\epsilon_k + \epsilon_{k+\pi}) f(\epsilon_{k+\pi}) [1-f(\epsilon_k)]\,,
\ee
with $\epsilon_k=-2[J+\mu \cos(k)]$ the single-particle energy of the diagonalised Hamiltonian and $f(\epsilon_k)=1/[\exp(\epsilon_k/ T)+1]$ the Fermi--Dirac occupation. 
As a consequence of the gaplessness, the QFI evaluated at ${q=0}$ vanishes, while it assumes its maximum value at ${q=\pi}$, i.e., for the staggered density. 
Using the fluctuation-dissipation theorem and evaluating the integral in Eq.~\eqref{eq:FQImChiIntro}, the corresponding quantum Fisher information density becomes 
\be
	f_Q = \frac 4 N \sum_k \tanh^2\left(\frac{\epsilon_k - \epsilon_{k+\pi}}{2 T}\right) f(\epsilon_{k+\pi}) [1-f(\epsilon_k)]\,.
\ee
This quantity is plotted in Fig.~\ref{fig:FQXY}(a) as a function of temperature and chemical potential, for $N=1000$. 
The entanglement witnessed by the QFI extends up to $ T\approx 0.91 J$, improving over other witnesses showing entanglement up to $ T\approx 0.7 J$ \cite{Guehne2006,Hide2007}. 
In contrast to Ref.~\cite{Guehne2006}, however, the QFI for the staggered magnetisation does not witness genuine multipartiteness of entanglement, but saturates at the value $f_Q=2$. 
(Notably, the QFI for other generators such as $b_l+b_l^\dagger$ does show divergent behaviour.) 
Towards the phase transition, the QFI for the staggered density decreases, consistent with the presence of the factorisation point. 
Despite this absence of divergences, the phase transition imprints its character onto the universal behaviour of the QFI and its derivatives, as explained in what follows.

At $T=0$ and $N=\infty$, one can use Eq.~\eqref{eq:QFIpure} to derive a particularly simple analytical formula for the QFI of the staggered density, 
\be
f_Q = \left\{ 
	\begin{array}{ll} 
	4 n & \mu \leq 0 \\
	4(1-n)& \mu > 0
	\end{array}
	\right.\,.
\ee
Here, we defined the mean particle density $n\equiv \braket{\hat{n}}$, which is $n = 1 - \mathrm{acos}(\mu/J)/\pi$ for $|\mu|<J$ and $0$ else. 
The quantum Fisher information density attains its maximal value of $f_Q= 2$ at $\mu=0$ and decreases towards the phase-transition point, where it vanishes. 
The above analytic expression allows us to obtain the scaling dimension of $f_Q$ by expanding it close to the critical point, $f_Q \propto \tilde{h}^{1/2}$, with $\tilde{h}=1-\mu/J$. 
Comparison with the scaling form Eq.~\eqref{eq:fQscalingwithh} gives the exact scaling exponent $\Delta_Q=-1$. 
Since $\Delta_Q<0$, $f_Q$ of the staggered magnetisation displays no divergent behaviour. 

Nevertheless, as seen in the inset of Fig.~\ref{fig:FQXY}(b), the non-analyticity of $f_Q$ at $\tilde{h}=0$ leads to a divergence in its derivatives. 
Considering the temperature as the most relevant perturbation, Eq.~\eqref{eq:fQscalingwithT}, we obtain the universal behaviour
\be
\frac{\partial f_Q}{\partial {\tilde{h}}}= T^{-\Delta_Q/z+1/(z\nu)}{\phi}_{Q3}(\tilde{h}T^{-1/(z\nu)})\,.
\ee 
That is, $\partial f_Q/\partial {\tilde{h}}$ has a peak that diverges as $T^{-\Delta_Q/z+1/(z\nu)}$ and the position of which lies at a line $\tilde h \sim T^{1/(z \nu)}$. 
Using the analytically known exponents $-\Delta_Q/z+1/(z\nu)=1/2$ and $1/(z \nu) =1$, an excellent scaling collapse is achieved up to rather large temperatures [main panel of Fig.~\ref{fig:FQXY}(b)]. 
We also performed fits to the numerical data for the peak height and peak position. These yield exponents of ${0.49}$ and ${1.02}$, respectively---even without high-precision data, we are able to reproduce the exact scaling exponents within two percent accuracy.


\subsection{Universal scaling in the infinite-range quantum Ising model}

\begin{figure}
\centering
\includegraphics[width = 0.7\columnwidth]{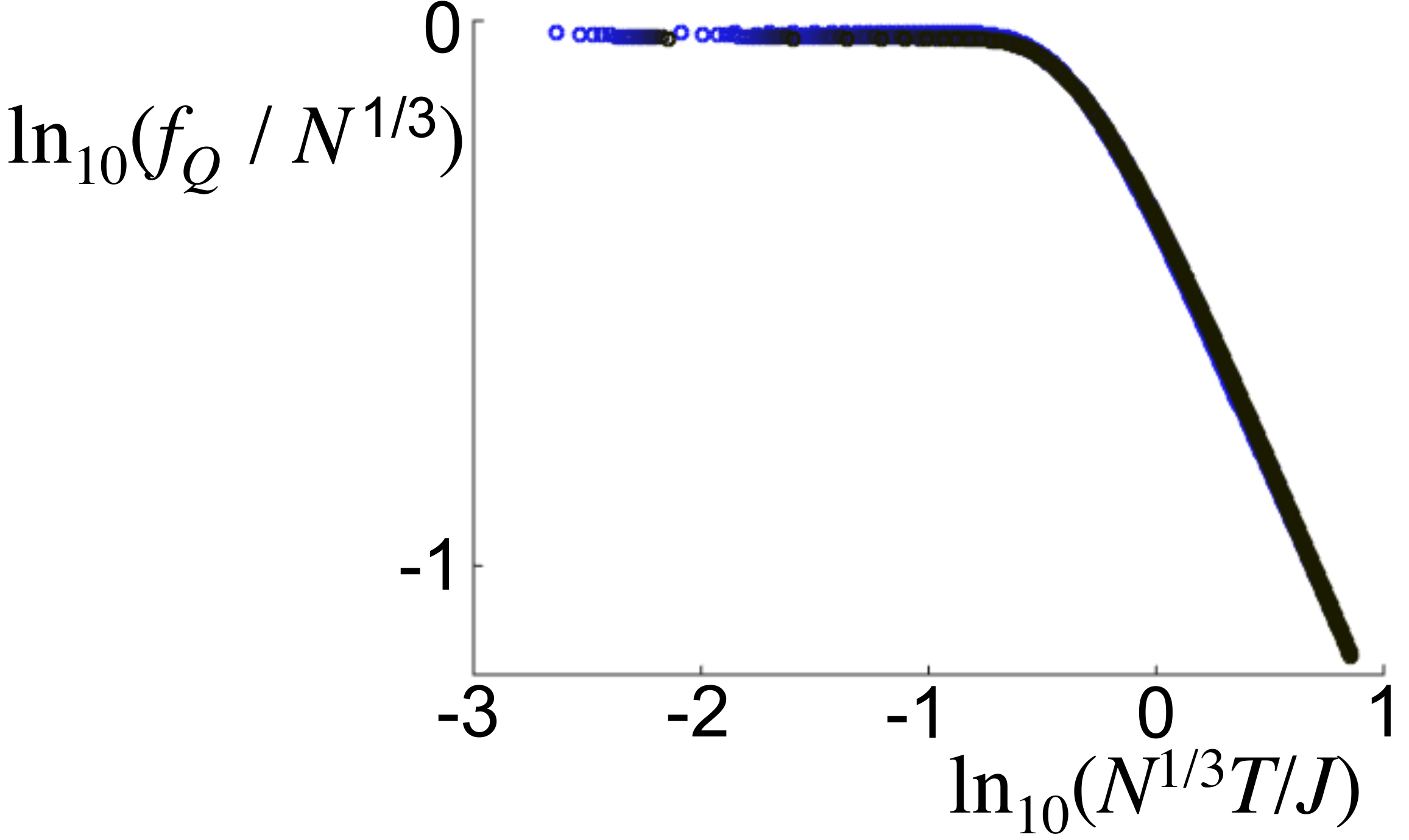}
\caption{(color online) Universal scaling of the quantum Fisher information density, $f_Q=F_Q/N$, calculated for the order parameter in the quantum Ising model with infinite-range interactions. 
Finite-temperature and finite-size scaling at criticality $\theta=\pi/4$. Bullets are from lighter to darker blue for $N=100,\dots,3000$. 
The universal scaling laws produce an excellent data collapse. The divergence $f_Q\sim N^{1/3}$ is much weaker than in the one-dimensional quantum Ising model, reflecting the more mean-field-like character of the infinite-range model. Nevertheless, it still implies a diverging multipartiteness of entanglement.}   
\label{fig:scaling_longrangeIsing}
\end{figure}

In the main text, we considered the infinite-range Ising model in a transverse field, main-text Eq.~\eqref{eq:Ising} with $\mathcal{J}_{lj}=1/N$, to illustrate the absence of features in the QFI at thermal transitions (see Fig.~\ref{fig:FQ_LRIsing}). 
Here, we provide its scaling analysis in the vicinity of its quantum phase transition.  

The infinite-range Ising model describes the universal properties above the upper critical dimension, with the longitudinal magnetization $\obs = \sum_{l=1}^N \sigma_l^x$ as the order parameter. 
In Fig.~\ref{fig:scaling_longrangeIsing}, we plot the QFI as a function of temperature for different system sizes at the system's quantum critical point, $\tilde h=0$. Using that $\Delta_Q=1/3$ and $z=1/3$~\cite{Botet1983} one obtains the scaling of the quantum Fisher information density $ f_Q \sim N^{1/3} \phi(T N^{1/3})$. From a scaling collapse, shown in Fig.~\ref{fig:scaling_longrangeIsing}, we determined numerically that $\Delta_Q = 0.34$ and $z=0.33$ which is in very good agreement with the exact solution.

 \section{Classical contribution to QFI and behaviour at thermal phase transitions}

In the main text, we study how well thermal states that lie close in a phase diagram are distinguishable, focusing entirely on quantum effects. 
More precisely, we ask how fast the probability distribution of a thermal state $\rho=\sum_\lambda p_\lambda \ket{\lambda}\bra{\lambda}$, with $p_\lambda=\exp(-E_\lambda/T)/Z$, changes due to the unitary transformation with $\ue^{i\delta\vartheta \obs}$. 
It is instructive to extend this analysis to any source of distinguishability, be it quantum or classical. 

Consider a canonical ensemble $\rho$ at temperature $T$ for a Hamiltonian $H$. We wish to compare this state with the equilibrium ensemble $\rho'$ at the same temperature but for $H' = H+{\cal J}{\delta\vartheta} \obs$, where $\delta\vartheta\ll1$ and where ${\cal J}$ represents a suitable energy scale. 
There are two sources that render $\rho$ and $\rho'$ distinguishable. 
The first is the one considered in the main text. It can be rephrased as an adiabatic evolution of the state by slowly changing the Hamiltonian from $H$ to $H'$. 
If the evolution is adiabatic and unitary, the populations $p_\lambda$ remain unchanged. Any difference between initial and final density operator is then solely due to a modification of the energy eigenstates. Taking a unit time step $t=1/{\cal J}$, the change amounts to $\ket{\lambda}\to\ue^{i\delta\vartheta \obs}\ket{\lambda}$. This is formally equivalent to the phase estimation scenario considered in this article. 

However, one may extend the scenario by including a sufficiently long waiting time after the transfer, such that the populations at the new Hamiltonian $H'$ can rethermalise with an external bath. This change of populations provides a second source of distinguishability between $\rho$ and $\rho'$, which to leading order remains disparate from the one described above.  
We can incorporate this effect by considering the full QFI \cite{BraunsteinCaves1994} 
\bea 
\tilde{F}_Q&=&F_C+F_Q\\
&=&\sum_\lambda \frac{(\partial p_\lambda/\partial\vartheta)^2}{p_\lambda} + 2\sum_{\lambda,\lambda^\prime} \frac{(p_\lambda-p_{\lambda^\prime})^2}{p_\lambda+p_{\lambda^\prime}} \left|\bra{\lambda}\obs\ket{\lambda^\prime}\right|^2 \nonumber \,. 
\eea
The second term, $F_Q$, the subject of this work, is the term that is relevant at a QPT at $T=0$, where the first term, $F_C$, vanishes identically. 

Both terms are related to different aspects of the susceptibility of $\rho$ towards the generator $\obs$. 
As demonstrated in Eq.~(2) of the main text, $F_Q$ is in one-to-one correspondence to the Kubo dynamic susceptibility at non-zero frequencies. In Kubo's linear response theory, a system is perturbed weakly in a time-dependent way, generating a unitary dynamics, but---to leading order---without changing the populations of energy levels \cite{JensenBookChapter3}. 

The change of probabilities, instead, is related to the static, isothermal susceptibility, as has already been pointed out within purely classical statistical physics \cite{ProkopenkoObst2011}. 
Here, we demonstrate this relation in a quantum statistical calculation. 
Consider a static perturbation, $H \to H' = H + {\delta\vartheta'}\obs$, with $\delta\vartheta'=T\delta\vartheta$ (i.e., we choose the energy scale ${\cal J}=T$). 
The expectation value of the observable $\obs$ is given by a derivative of the free entropy $\ln Z$, with $Z=\sum_\lambda \exp(-E_\lambda/T)$ the partition function, 
\be
\braket{\obs}=T \frac{\partial \ln Z}{\partial \vartheta'} = - \sum_\lambda p_\lambda \frac{\partial E_\lambda}{ \partial\vartheta'}\,,
\ee
and the corresponding static, isothermal susceptibility is 
\be
\chi^T \equiv \frac{\partial\braket{\obs}}{\partial \vartheta'} = - \sum_\lambda  \frac{\partial p_\lambda}{ \partial\vartheta'} \frac{\partial E_\lambda}{ \partial\vartheta'}  - \sum_{\lambda}   p_\lambda \frac{\partial^2 E_\lambda}{ (\partial\vartheta')^2} \,.
\ee 
The derivative of $p_\lambda$ reads  
\be
\label{eq:derivativeofplambda}
\frac{\partial p_\lambda}{ \partial\vartheta'}=-\frac {p_\lambda} T \left( \frac{\partial E_\lambda}{ \partial\vartheta'}-\sum_{\lambda'} \frac{\partial E_{\lambda'}}{ \partial\vartheta'}\right)
\ee
One can use perturbation theory to evaluate the change of energy levels following the static perturbation. Working in a basis where degenerate eigenlevels $\ket{\lambda}$ are prediagonalised with respect to the perturbation operator $\obs$, we obtain    
\bea
\chi^T &=&  \frac 1 T \left( \sum_\lambda p_\lambda \obs_{\lambda,\lambda}^2 - \braket{\obs}^2\right) + \sum_{\lambda,\lambda'}^{E_\lambda\neq E_\lambda'} \frac{|\obs_{\lambda,\lambda'}|^2 (p_\lambda-p_\lambda')}{E_{\lambda'}-E_\lambda} \nonumber \\
&=& \chi_{\mathrm{el}} + \chi_{\mathrm{vV}}\,.
\eea
Here, we defined the abbreviation $\obs_{\lambda,\lambda'}\equiv \bra{\lambda}\obs\ket{\lambda'}$. 
The first term is the elastic contribution, or classical Curie contribution, arising from a change of populations. It vanishes in a pure state, such as the quantum ground state at $T=0$. 
The second term is the quantum-mechanical van-Vleck correction, arising from matrix elements between different energy levels. It is connected smoothly to the real part of the Kubo dynamic susceptibility at non-zero frequencies and vanishes in a purely classical theory. 
Importantly, the elastic peak is a singular contribution at exactly $\omega=0$, and does thus not influence $F_Q$. Since it is this elastic peak that diverges at thermal phase transitions, these do not appear as particular features in $F_Q$. 

We can connect the elastic peak to the classical contribution to the Fisher information \cite{ProkopenkoObst2011}. Equation~\eqref{eq:derivativeofplambda} yields $F_C=\sum_\lambda \frac{(\partial p_\lambda/\partial\vartheta)^2}{p_\lambda}= \chi_{\mathrm{el}}/T$. 
Similarly to $\chi^T$, $\tilde{F}_Q$ thus has two parts: $F_Q$, the subject of this article, related to entanglement and divergent at QPTs; 
and $F_C$, of non-unitary, classical origin and divergent at thermal phase transitions. 
Both are related to the order parameter's susceptibility. Remarkably, by choosing the susceptibility at vanishing or non-vanishing frequencies, we can extract the purely classical or the purely quantum-mechanical contribution to the quantum Fisher information. 


 \section{Sum rule}
 
Our central relation Eq.~(2) allows us to derive a Thomas-Reiche-Kuhn-type \cite{Thomas1925,Reiche1925,Kuhn1925} sum rule that may be useful to bound the QFI. 
Since $\tanh\left(\frac{\omega}{2T}\right)\leq\frac{\omega}{2T}$, we have 
\be
	F_Q(T) \leq \frac{1}{T}  \frac{2}{\pi}\int_{0}^{\infty} \ud \omega \,\omega\, {\chi}''(\omega,T)\,.
\ee 
Due to the Kramers--Kronig relations, we can now take advantage of a superconvergence theorem \cite{Lucarini2003}. 
This theorem states that two complex functions $g$ and $f$, for which 
$g(x) =  P \int_0^\infty \ud \tilde{x}\, \frac{f(\tilde{x})}{x^2-\tilde{x}^2}$ holds, are related via 
$
\int_0^\infty \ud x \, f(x) = \lim_{x\to\infty} \left(x^2 g(x)\right)\,,  
$
with $P$ the principal part of the integral. 
By setting $f(x)=-\frac 2 \pi x \chi''(x)$ and $g(x)=\chi'(x)\equiv\Re(\chi(x))$, one obtains
\be
	\label{eq:FQsumRule}
	F_Q(T) \leq - \frac 1 T \lim_{\omega\to\infty}(\omega^2 \chi'(\omega,T))\,, 
\ee
a Thomas-Reiche-Kuhn-type sum rule that upper bounds the QFI.  
Here, as in Eq.~\eqref{eq:FQImChiIntro} of the main text, the limit to infinity is to be understood as a frequency range much larger than all energy scales of the considered degrees of freedom (but smaller than energy scales at which resonances to additional degrees of freedom appear, the entanglement between which, however, could be analysed analogously).   
It is interesting to note that, if we take the classical limit by sending $T\to \infty$, the bound \eqref{eq:FQsumRule} becomes an equality relating the QFI to the high-frequency limit of $\chi'(\omega)$.\\


%




\begin{thebibliography}{0}%
\makeatletter
\providecommand \@ifxundefined [1]{%
 \@ifx{#1\undefined}
}%
\providecommand \@ifnum [1]{%
 \ifnum #1\expandafter \@firstoftwo
 \else \expandafter \@secondoftwo
 \fi
}%
\providecommand \@ifx [1]{%
 \ifx #1\expandafter \@firstoftwo
 \else \expandafter \@secondoftwo
 \fi
}%
\providecommand \natexlab [1]{#1}%
\providecommand \enquote  [1]{``#1''}%
\providecommand \bibnamefont  [1]{#1}%
\providecommand \bibfnamefont [1]{#1}%
\providecommand \citenamefont [1]{#1}%
\providecommand \href@noop [0]{\@secondoftwo}%
\providecommand \href [0]{\begingroup \@sanitize@url \@href}%
\providecommand \@href[1]{\@@startlink{#1}\@@href}%
\providecommand \@@href[1]{\endgroup#1\@@endlink}%
\providecommand \@sanitize@url [0]{\catcode `\\12\catcode `\$12\catcode
  `\&12\catcode `\#12\catcode `\^12\catcode `\_12\catcode `\%12\relax}%
\providecommand \@@startlink[1]{}%
\providecommand \@@endlink[0]{}%
\providecommand \url  [0]{\begingroup\@sanitize@url \@url }%
\providecommand \@url [1]{\endgroup\@href {#1}{\urlprefix }}%
\providecommand \urlprefix  [0]{URL }%
\providecommand \Eprint [0]{\href }%
\providecommand \doibase [0]{http://dx.doi.org/}%
\providecommand \selectlanguage [0]{\@gobble}%
\providecommand \bibinfo  [0]{\@secondoftwo}%
\providecommand \bibfield  [0]{\@secondoftwo}%
\providecommand \translation [1]{[#1]}%
\providecommand \BibitemOpen [0]{}%
\providecommand \bibitemStop [0]{}%
\providecommand \bibitemNoStop [0]{.\EOS\space}%
\providecommand \EOS [0]{\spacefactor3000\relax}%
\providecommand \BibitemShut  [1]{\csname bibitem#1\endcsname}%
\let\auto@bib@innerbib\@empty
\end{thebibliography}%


\begin{thebibliography}{61}%
\makeatletter
\providecommand \@ifxundefined [1]{%
 \@ifx{#1\undefined}
}%
\providecommand \@ifnum [1]{%
 \ifnum #1\expandafter \@firstoftwo
 \else \expandafter \@secondoftwo
 \fi
}%
\providecommand \@ifx [1]{%
 \ifx #1\expandafter \@firstoftwo
 \else \expandafter \@secondoftwo
 \fi
}%
\providecommand \natexlab [1]{#1}%
\providecommand \enquote  [1]{``#1''}%
\providecommand \bibnamefont  [1]{#1}%
\providecommand \bibfnamefont [1]{#1}%
\providecommand \citenamefont [1]{#1}%
\providecommand \href@noop [0]{\@secondoftwo}%
\providecommand \href [0]{\begingroup \@sanitize@url \@href}%
\providecommand \@href[1]{\@@startlink{#1}\@@href}%
\providecommand \@@href[1]{\endgroup#1\@@endlink}%
\providecommand \@sanitize@url [0]{\catcode `\\12\catcode `\$12\catcode
  `\&12\catcode `\#12\catcode `\^12\catcode `\_12\catcode `\%12\relax}%
\providecommand \@@startlink[1]{}%
\providecommand \@@endlink[0]{}%
\providecommand \url  [0]{\begingroup\@sanitize@url \@url }%
\providecommand \@url [1]{\endgroup\@href {#1}{\urlprefix }}%
\providecommand \urlprefix  [0]{URL }%
\providecommand \Eprint [0]{\href }%
\providecommand \doibase [0]{http://dx.doi.org/}%
\providecommand \selectlanguage [0]{\@gobble}%
\providecommand \bibinfo  [0]{\@secondoftwo}%
\providecommand \bibfield  [0]{\@secondoftwo}%
\providecommand \translation [1]{[#1]}%
\providecommand \BibitemOpen [0]{}%
\providecommand \bibitemStop [0]{}%
\providecommand \bibitemNoStop [0]{.\EOS\space}%
\providecommand \EOS [0]{\spacefactor3000\relax}%
\providecommand \BibitemShut  [1]{\csname bibitem#1\endcsname}%
\let\auto@bib@innerbib\@empty
\bibitem [{\citenamefont {Balents}(2010)}]{Balents2010}%
  \BibitemOpen
  \bibfield  {author} {\bibinfo {author} {\bibfnamefont {L.}~\bibnamefont
  {Balents}},\ }\href@noop {} {\bibfield  {journal} {\bibinfo  {journal}
  {Nature}\ }\textbf {\bibinfo {volume} {464}},\ \bibinfo {pages} {199}
  (\bibinfo {year} {2010})}\BibitemShut {NoStop}%
\bibitem [{\citenamefont {Nandkishore}\ and\ \citenamefont
  {Huse}(2015)}]{Nandkishore2014}%
  \BibitemOpen
  \bibfield  {author} {\bibinfo {author} {\bibfnamefont {R.}~\bibnamefont
  {Nandkishore}}\ and\ \bibinfo {author} {\bibfnamefont {D.~A.}\ \bibnamefont
  {Huse}},\ }\href@noop {} {\bibfield  {journal} {\bibinfo  {journal} {Annu.
  Rev. Condens. Matter Phys.}\ }\textbf {\bibinfo {volume} {6}},\ \bibinfo
  {pages} {15} (\bibinfo {year} {2015})}\BibitemShut {NoStop}%
\bibitem [{\citenamefont {Smith}\ \emph {et~al.}(2015)\citenamefont {Smith},
  \citenamefont {Lee}, \citenamefont {Richerme}, \citenamefont {Neyenhuis},
  \citenamefont {Hess}, \citenamefont {Hauke}, \citenamefont {Heyl},
  \citenamefont {Huse},\ and\ \citenamefont {Monroe}}]{Smith2015}%
  \BibitemOpen
  \bibfield  {author} {\bibinfo {author} {\bibfnamefont {J.}~\bibnamefont
  {Smith}}, \bibinfo {author} {\bibfnamefont {A.}~\bibnamefont {Lee}}, \bibinfo
  {author} {\bibfnamefont {P.}~\bibnamefont {Richerme}}, \bibinfo {author}
  {\bibfnamefont {B.}~\bibnamefont {Neyenhuis}}, \bibinfo {author}
  {\bibfnamefont {P.~W.}\ \bibnamefont {Hess}}, \bibinfo {author}
  {\bibfnamefont {P.}~\bibnamefont {Hauke}}, \bibinfo {author} {\bibfnamefont
  {M.}~\bibnamefont {Heyl}}, \bibinfo {author} {\bibfnamefont {D.}~\bibnamefont
  {Huse}}, \ and\ \bibinfo {author} {\bibfnamefont {C.}~\bibnamefont
  {Monroe}},\ }\href@noop {} {\bibfield  {journal} {\bibinfo  {journal}
  {arXiv:1508.07026 [quant-ph]}\ } (\bibinfo {year} {2015})}\BibitemShut
  {NoStop}%
\bibitem [{\citenamefont {Nishioka}\ and\ \citenamefont
  {Takayanagi}(2007)}]{Nishioka2007}%
  \BibitemOpen
  \bibfield  {author} {\bibinfo {author} {\bibfnamefont {T.}~\bibnamefont
  {Nishioka}}\ and\ \bibinfo {author} {\bibfnamefont {T.}~\bibnamefont
  {Takayanagi}},\ }\href@noop {} {\bibfield  {journal} {\bibinfo  {journal}
  {JHEP}\ }\textbf {\bibinfo {volume} {0701}},\ \bibinfo {pages} {090}
  (\bibinfo {year} {2007})}\BibitemShut {NoStop}%
\bibitem [{\citenamefont {Klebanov}\ \emph {et~al.}(2008)\citenamefont
  {Klebanov}, \citenamefont {Kutasov},\ and\ \citenamefont
  {Murugan}}]{Klebanov2008}%
  \BibitemOpen
  \bibfield  {author} {\bibinfo {author} {\bibfnamefont {I.~R.}\ \bibnamefont
  {Klebanov}}, \bibinfo {author} {\bibfnamefont {D.}~\bibnamefont {Kutasov}}, \
  and\ \bibinfo {author} {\bibfnamefont {A.}~\bibnamefont {Murugan}},\
  }\href@noop {} {\bibfield  {journal} {\bibinfo  {journal} {Nucl. Phys. B}\
  }\textbf {\bibinfo {volume} {796}},\ \bibinfo {pages} {274} (\bibinfo {year}
  {2008})}\BibitemShut {NoStop}%
\bibitem [{\citenamefont {Escher}\ \emph {et~al.}(2011)\citenamefont {Escher},
  \citenamefont {de~Matos~Filho},\ and\ \citenamefont
  {Davidovich}}]{Escher2011}%
  \BibitemOpen
  \bibfield  {author} {\bibinfo {author} {\bibfnamefont {B.~M.}\ \bibnamefont
  {Escher}}, \bibinfo {author} {\bibfnamefont {R.~L.}\ \bibnamefont
  {de~Matos~Filho}}, \ and\ \bibinfo {author} {\bibfnamefont {L.}~\bibnamefont
  {Davidovich}},\ }\href@noop {} {\bibfield  {journal} {\bibinfo  {journal}
  {Nature Phys.}\ }\textbf {\bibinfo {volume} {7}},\ \bibinfo {pages} {406}
  (\bibinfo {year} {2011})}\BibitemShut {NoStop}%
\bibitem [{\citenamefont {Pezz{\'e}}\ and\ \citenamefont
  {Smerzi}(2014)}]{Pezze2014}%
  \BibitemOpen
  \bibfield  {author} {\bibinfo {author} {\bibfnamefont {L.}~\bibnamefont
  {Pezz{\'e}}}\ and\ \bibinfo {author} {\bibfnamefont {A.}~\bibnamefont
  {Smerzi}},\ }in\ \href@noop {} {\emph {\bibinfo {booktitle} {''Atom
  Interferometry''}}},\ \bibinfo {series and number} {Proceedings of the
  International School of Physics ''Enrico Fermi'', Course 188, Varenna},\
  \bibinfo {editor} {edited by\ \bibinfo {editor} {\bibfnamefont
  {G.}~\bibnamefont {Tino}}\ and\ \bibinfo {editor} {\bibfnamefont
  {M.}~\bibnamefont {Kasevich}}}\ (\bibinfo  {publisher} {IOS Press,
  Amsterdam},\ \bibinfo {year} {2014})\ p.\ \bibinfo {pages} {691}\BibitemShut
  {NoStop}%
\bibitem [{\citenamefont {Ladd}\ \emph {et~al.}(2010)\citenamefont {Ladd},
  \citenamefont {Jelezko}, \citenamefont {Laflamme}, \citenamefont {Nakamura},
  \citenamefont {Monroe},\ and\ \citenamefont {O'Brien}}]{Ladd2010}%
  \BibitemOpen
  \bibfield  {author} {\bibinfo {author} {\bibfnamefont {T.~D.}\ \bibnamefont
  {Ladd}}, \bibinfo {author} {\bibfnamefont {F.}~\bibnamefont {Jelezko}},
  \bibinfo {author} {\bibfnamefont {R.}~\bibnamefont {Laflamme}}, \bibinfo
  {author} {\bibfnamefont {Y.}~\bibnamefont {Nakamura}}, \bibinfo {author}
  {\bibfnamefont {C.}~\bibnamefont {Monroe}}, \ and\ \bibinfo {author}
  {\bibfnamefont {J.~L.}\ \bibnamefont {O'Brien}},\ }\href@noop {} {\bibfield
  {journal} {\bibinfo  {journal} {Nature}\ }\textbf {\bibinfo {volume} {464}},\
  \bibinfo {pages} {45} (\bibinfo {year} {2010})}\BibitemShut {NoStop}%
\bibitem [{\citenamefont {Cirac}\ and\ \citenamefont
  {Zoller}(2012)}]{Cirac2012}%
  \BibitemOpen
  \bibfield  {author} {\bibinfo {author} {\bibfnamefont {J.~I.}\ \bibnamefont
  {Cirac}}\ and\ \bibinfo {author} {\bibfnamefont {P.}~\bibnamefont {Zoller}},\
  }\href@noop {} {\bibfield  {journal} {\bibinfo  {journal} {Nature Phys.}\
  }\textbf {\bibinfo {volume} {8}},\ \bibinfo {pages} {264} (\bibinfo {year}
  {2012})}\BibitemShut {NoStop}%
\bibitem [{\citenamefont {Hauke}\ \emph {et~al.}(2012)\citenamefont {Hauke},
  \citenamefont {Cucchietti}, \citenamefont {Tagliacozzo}, \citenamefont
  {Deutsch},\ and\ \citenamefont {Lewenstein}}]{Hauke2011d}%
  \BibitemOpen
  \bibfield  {author} {\bibinfo {author} {\bibfnamefont {P.}~\bibnamefont
  {Hauke}}, \bibinfo {author} {\bibfnamefont {F.~M.}\ \bibnamefont
  {Cucchietti}}, \bibinfo {author} {\bibfnamefont {L.}~\bibnamefont
  {Tagliacozzo}}, \bibinfo {author} {\bibfnamefont {I.}~\bibnamefont
  {Deutsch}}, \ and\ \bibinfo {author} {\bibfnamefont {M.}~\bibnamefont
  {Lewenstein}},\ }\href@noop {} {\bibfield  {journal} {\bibinfo  {journal}
  {Rep. Prog. Phys.}\ }\textbf {\bibinfo {volume} {75}},\ \bibinfo {pages}
  {082401} (\bibinfo {year} {2012})}\BibitemShut {NoStop}%
\bibitem [{\citenamefont {Georgescu}\ \emph {et~al.}(2014)\citenamefont
  {Georgescu}, \citenamefont {Ashhab},\ and\ \citenamefont
  {Nori}}]{Georgescu2014}%
  \BibitemOpen
  \bibfield  {author} {\bibinfo {author} {\bibfnamefont {I.~M.}\ \bibnamefont
  {Georgescu}}, \bibinfo {author} {\bibfnamefont {S.}~\bibnamefont {Ashhab}}, \
  and\ \bibinfo {author} {\bibfnamefont {F.}~\bibnamefont {Nori}},\ }\href@noop
  {} {\bibfield  {journal} {\bibinfo  {journal} {Rev. Mod. Phys.}\ }\textbf
  {\bibinfo {volume} {86}},\ \bibinfo {pages} {154} (\bibinfo {year}
  {2014})}\BibitemShut {NoStop}%
\bibitem [{\citenamefont {Kim}\ \emph {et~al.}(2010)\citenamefont {Kim},
  \citenamefont {Chang}, \citenamefont {Korenblit}, \citenamefont {Islam},
  \citenamefont {Edwards}, \citenamefont {Freericks}, \citenamefont {Lin},
  \citenamefont {Duan},\ and\ \citenamefont {Monroe}}]{Kim2010}%
  \BibitemOpen
  \bibfield  {author} {\bibinfo {author} {\bibfnamefont {K.}~\bibnamefont
  {Kim}}, \bibinfo {author} {\bibfnamefont {M.-S.}\ \bibnamefont {Chang}},
  \bibinfo {author} {\bibfnamefont {S.}~\bibnamefont {Korenblit}}, \bibinfo
  {author} {\bibfnamefont {R.}~\bibnamefont {Islam}}, \bibinfo {author}
  {\bibfnamefont {E.~E.}\ \bibnamefont {Edwards}}, \bibinfo {author}
  {\bibfnamefont {J.~K.}\ \bibnamefont {Freericks}}, \bibinfo {author}
  {\bibfnamefont {G.-D.}\ \bibnamefont {Lin}}, \bibinfo {author} {\bibfnamefont
  {L.-M.}\ \bibnamefont {Duan}}, \ and\ \bibinfo {author} {\bibfnamefont
  {C.}~\bibnamefont {Monroe}},\ }\href@noop {} {\bibfield  {journal} {\bibinfo
  {journal} {Nature}\ }\textbf {\bibinfo {volume} {465}},\ \bibinfo {pages}
  {590} (\bibinfo {year} {2010})}\BibitemShut {NoStop}%
\bibitem [{\citenamefont {Jurcevic}\ \emph {et~al.}(2014)\citenamefont
  {Jurcevic}, \citenamefont {Lanyon}, \citenamefont {Hauke}, \citenamefont
  {Hempel}, \citenamefont {Zoller}, \citenamefont {Blatt},\ and\ \citenamefont
  {Roos}}]{Jurcevic2014}%
  \BibitemOpen
  \bibfield  {author} {\bibinfo {author} {\bibfnamefont {P.}~\bibnamefont
  {Jurcevic}}, \bibinfo {author} {\bibfnamefont {B.~P.}\ \bibnamefont
  {Lanyon}}, \bibinfo {author} {\bibfnamefont {P.}~\bibnamefont {Hauke}},
  \bibinfo {author} {\bibfnamefont {C.}~\bibnamefont {Hempel}}, \bibinfo
  {author} {\bibfnamefont {P.}~\bibnamefont {Zoller}}, \bibinfo {author}
  {\bibfnamefont {R.}~\bibnamefont {Blatt}}, \ and\ \bibinfo {author}
  {\bibfnamefont {C.~F.}\ \bibnamefont {Roos}},\ }\href@noop {} {\bibfield
  {journal} {\bibinfo  {journal} {Nature}\ }\textbf {\bibinfo {volume} {511}},\
  \bibinfo {pages} {202} (\bibinfo {year} {2014})}\BibitemShut {NoStop}%
\bibitem [{\citenamefont {{Lanting et al.}}(2014)}]{Lanting2014}%
  \BibitemOpen
  \bibfield  {author} {\bibinfo {author} {\bibfnamefont {T.}~\bibnamefont
  {{Lanting et al.}}},\ }\href@noop {} {\bibfield  {journal} {\bibinfo
  {journal} {Phys. Rev. X}\ }\textbf {\bibinfo {volume} {4}},\ \bibinfo {pages}
  {021041} (\bibinfo {year} {2014})}\BibitemShut {NoStop}%
\bibitem [{\citenamefont {Fukuhara}\ \emph {et~al.}(2015)\citenamefont
  {Fukuhara}, \citenamefont {Hild}, \citenamefont {Zeiher}, \citenamefont
  {Schau{\ss}}, \citenamefont {Bloch}, \citenamefont {Endres},\ and\
  \citenamefont {Gross}}]{Fukuhara2015}%
  \BibitemOpen
  \bibfield  {author} {\bibinfo {author} {\bibfnamefont {T.}~\bibnamefont
  {Fukuhara}}, \bibinfo {author} {\bibfnamefont {S.}~\bibnamefont {Hild}},
  \bibinfo {author} {\bibfnamefont {J.}~\bibnamefont {Zeiher}}, \bibinfo
  {author} {\bibfnamefont {P.}~\bibnamefont {Schau{\ss}}}, \bibinfo {author}
  {\bibfnamefont {I.}~\bibnamefont {Bloch}}, \bibinfo {author} {\bibfnamefont
  {M.}~\bibnamefont {Endres}}, \ and\ \bibinfo {author} {\bibfnamefont
  {C.}~\bibnamefont {Gross}},\ }\href@noop {} {\bibfield  {journal} {\bibinfo
  {journal} {Phys. Rev. Lett.}\ }\textbf {\bibinfo {volume} {115}},\ \bibinfo
  {pages} {035302} (\bibinfo {year} {2015})}\BibitemShut {NoStop}%
\bibitem [{\citenamefont {Daley}\ \emph {et~al.}(2012)\citenamefont {Daley},
  \citenamefont {Pichler}, \citenamefont {Schachenmayer},\ and\ \citenamefont
  {Zoller}}]{Daley2012}%
  \BibitemOpen
  \bibfield  {author} {\bibinfo {author} {\bibfnamefont {A.~J.}\ \bibnamefont
  {Daley}}, \bibinfo {author} {\bibfnamefont {H.}~\bibnamefont {Pichler}},
  \bibinfo {author} {\bibfnamefont {J.}~\bibnamefont {Schachenmayer}}, \ and\
  \bibinfo {author} {\bibfnamefont {P.}~\bibnamefont {Zoller}},\ }\href@noop {}
  {\bibfield  {journal} {\bibinfo  {journal} {Phys. Rev. Lett.}\ }\textbf
  {\bibinfo {volume} {109}},\ \bibinfo {pages} {020505} (\bibinfo {year}
  {2012})}\BibitemShut {NoStop}%
\bibitem [{\citenamefont {Islam}\ \emph {et~al.}(2015)\citenamefont {Islam},
  \citenamefont {Ma}, \citenamefont {Preiss}, \citenamefont {Tai},
  \citenamefont {Lukin}, \citenamefont {Rispoli},\ and\ \citenamefont
  {Greiner}}]{Islam2015}%
  \BibitemOpen
  \bibfield  {author} {\bibinfo {author} {\bibfnamefont {R.}~\bibnamefont
  {Islam}}, \bibinfo {author} {\bibfnamefont {R.}~\bibnamefont {Ma}}, \bibinfo
  {author} {\bibfnamefont {P.~M.}\ \bibnamefont {Preiss}}, \bibinfo {author}
  {\bibfnamefont {M.~E.}\ \bibnamefont {Tai}}, \bibinfo {author} {\bibfnamefont
  {A.}~\bibnamefont {Lukin}}, \bibinfo {author} {\bibfnamefont
  {M.}~\bibnamefont {Rispoli}}, \ and\ \bibinfo {author} {\bibfnamefont
  {M.}~\bibnamefont {Greiner}},\ }\href@noop {} {\bibfield  {journal} {\bibinfo
   {journal} {arXiv:1509.01160 [cond-mat.quant-gas]}\ } (\bibinfo {year}
  {2015})}\BibitemShut {NoStop}%
\bibitem [{\citenamefont {Hyllus}\ \emph {et~al.}(2012)\citenamefont {Hyllus},
  \citenamefont {Laskowski}, \citenamefont {Krischek}, \citenamefont
  {Schwemmer}, \citenamefont {Wieczorek}, \citenamefont {Weinfurter},
  \citenamefont {Pezz{\'e}},\ and\ \citenamefont {Smerzi}}]{Hyllus2012}%
  \BibitemOpen
  \bibfield  {author} {\bibinfo {author} {\bibfnamefont {P.}~\bibnamefont
  {Hyllus}}, \bibinfo {author} {\bibfnamefont {W.}~\bibnamefont {Laskowski}},
  \bibinfo {author} {\bibfnamefont {R.}~\bibnamefont {Krischek}}, \bibinfo
  {author} {\bibfnamefont {C.}~\bibnamefont {Schwemmer}}, \bibinfo {author}
  {\bibfnamefont {W.}~\bibnamefont {Wieczorek}}, \bibinfo {author}
  {\bibfnamefont {H.}~\bibnamefont {Weinfurter}}, \bibinfo {author}
  {\bibfnamefont {L.}~\bibnamefont {Pezz{\'e}}}, \ and\ \bibinfo {author}
  {\bibfnamefont {A.}~\bibnamefont {Smerzi}},\ }\href@noop {} {\bibfield
  {journal} {\bibinfo  {journal} {Phys. Rev. A}\ }\textbf {\bibinfo {volume}
  {85}},\ \bibinfo {pages} {022321} (\bibinfo {year} {2012})}\BibitemShut
  {NoStop}%
\bibitem [{\citenamefont {T{\'o}th}(2012)}]{Toth2012}%
  \BibitemOpen
  \bibfield  {author} {\bibinfo {author} {\bibfnamefont {G.}~\bibnamefont
  {T{\'o}th}},\ }\href@noop {} {\bibfield  {journal} {\bibinfo  {journal}
  {Phys. Rev. A}\ }\textbf {\bibinfo {volume} {85}},\ \bibinfo {pages} {022322}
  (\bibinfo {year} {2012})}\BibitemShut {NoStop}%
\bibitem [{\citenamefont {Strobel}\ \emph {et~al.}(2014)\citenamefont
  {Strobel}, \citenamefont {Muessel}, \citenamefont {Linnemann}, \citenamefont
  {Zibold}, \citenamefont {Hume}, \citenamefont {Pezz{\'e}}, \citenamefont
  {Smerzi},\ and\ \citenamefont {Oberthaler}}]{Strobel2014}%
  \BibitemOpen
  \bibfield  {author} {\bibinfo {author} {\bibfnamefont {H.}~\bibnamefont
  {Strobel}}, \bibinfo {author} {\bibfnamefont {W.}~\bibnamefont {Muessel}},
  \bibinfo {author} {\bibfnamefont {D.}~\bibnamefont {Linnemann}}, \bibinfo
  {author} {\bibfnamefont {T.}~\bibnamefont {Zibold}}, \bibinfo {author}
  {\bibfnamefont {D.~B.}\ \bibnamefont {Hume}}, \bibinfo {author}
  {\bibfnamefont {L.}~\bibnamefont {Pezz{\'e}}}, \bibinfo {author}
  {\bibfnamefont {A.}~\bibnamefont {Smerzi}}, \ and\ \bibinfo {author}
  {\bibfnamefont {M.~K.}\ \bibnamefont {Oberthaler}},\ }\href@noop {}
  {\bibfield  {journal} {\bibinfo  {journal} {Science}\ }\textbf {\bibinfo
  {volume} {345}},\ \bibinfo {pages} {424} (\bibinfo {year}
  {2014})}\BibitemShut {NoStop}%
\bibitem [{\citenamefont {Simon}\ \emph {et~al.}(2011)\citenamefont {Simon},
  \citenamefont {Bakr}, \citenamefont {Ma}, \citenamefont {Tai}, \citenamefont
  {Preiss},\ and\ \citenamefont {Greiner}}]{Simon2011}%
  \BibitemOpen
  \bibfield  {author} {\bibinfo {author} {\bibfnamefont {J.}~\bibnamefont
  {Simon}}, \bibinfo {author} {\bibfnamefont {W.~S.}\ \bibnamefont {Bakr}},
  \bibinfo {author} {\bibfnamefont {R.}~\bibnamefont {Ma}}, \bibinfo {author}
  {\bibfnamefont {M.~E.}\ \bibnamefont {Tai}}, \bibinfo {author} {\bibfnamefont
  {P.~M.}\ \bibnamefont {Preiss}}, \ and\ \bibinfo {author} {\bibfnamefont
  {M.}~\bibnamefont {Greiner}},\ }\href@noop {} {\bibfield  {journal} {\bibinfo
   {journal} {Nature}\ }\textbf {\bibinfo {volume} {472}},\ \bibinfo {pages}
  {307} (\bibinfo {year} {2011})}\BibitemShut {NoStop}%
\bibitem [{\citenamefont {Ma}\ and\ \citenamefont {Wang}(2009)}]{Ma2009}%
  \BibitemOpen
  \bibfield  {author} {\bibinfo {author} {\bibfnamefont {J.}~\bibnamefont
  {Ma}}\ and\ \bibinfo {author} {\bibfnamefont {X.}~\bibnamefont {Wang}},\
  }\href@noop {} {\bibfield  {journal} {\bibinfo  {journal} {Phys. Rev. A}\
  }\textbf {\bibinfo {volume} {80}},\ \bibinfo {pages} {012318} (\bibinfo
  {year} {2009})}\BibitemShut {NoStop}%
\bibitem [{\citenamefont {Liu}\ \emph {et~al.}(2013)\citenamefont {Liu},
  \citenamefont {Ma},\ and\ \citenamefont {Wang}}]{Liu2013QuantumFisher}%
  \BibitemOpen
  \bibfield  {author} {\bibinfo {author} {\bibfnamefont {W.-F.}\ \bibnamefont
  {Liu}}, \bibinfo {author} {\bibfnamefont {J.}~\bibnamefont {Ma}}, \ and\
  \bibinfo {author} {\bibfnamefont {X.}~\bibnamefont {Wang}},\ }\href@noop {}
  {\bibfield  {journal} {\bibinfo  {journal} {J. Phys. A: Math. Theor.}\
  }\textbf {\bibinfo {volume} {46}},\ \bibinfo {pages} {045302} (\bibinfo
  {year} {2013})}\BibitemShut {NoStop}%
\bibitem [{\citenamefont {Wang}\ \emph {et~al.}(2014)\citenamefont {Wang},
  \citenamefont {Wu}, \citenamefont {Yang}, \citenamefont {Jin}, \citenamefont
  {Lambert},\ and\ \citenamefont {Nori}}]{Wang2014}%
  \BibitemOpen
  \bibfield  {author} {\bibinfo {author} {\bibfnamefont {T.-L.}\ \bibnamefont
  {Wang}}, \bibinfo {author} {\bibfnamefont {L.-N.}\ \bibnamefont {Wu}},
  \bibinfo {author} {\bibfnamefont {W.}~\bibnamefont {Yang}}, \bibinfo {author}
  {\bibfnamefont {G.-R.}\ \bibnamefont {Jin}}, \bibinfo {author} {\bibfnamefont
  {N.}~\bibnamefont {Lambert}}, \ and\ \bibinfo {author} {\bibfnamefont
  {F.}~\bibnamefont {Nori}},\ }\href@noop {} {\bibfield  {journal} {\bibinfo
  {journal} {New J. Phys.}\ }\textbf {\bibinfo {volume} {16}},\ \bibinfo
  {pages} {063039} (\bibinfo {year} {2014})}\BibitemShut {NoStop}%
\bibitem [{\citenamefont {Zheng}\ \emph {et~al.}(2015)\citenamefont {Zheng},
  \citenamefont {Yao},\ and\ \citenamefont {Xu}}]{Zheng2015}%
  \BibitemOpen
  \bibfield  {author} {\bibinfo {author} {\bibfnamefont {Q.}~\bibnamefont
  {Zheng}}, \bibinfo {author} {\bibfnamefont {Y.}~\bibnamefont {Yao}}, \ and\
  \bibinfo {author} {\bibfnamefont {X.-W.}\ \bibnamefont {Xu}},\ }\href@noop {}
  {\bibfield  {journal} {\bibinfo  {journal} {Commun. Theor. Phys.}\ }\textbf
  {\bibinfo {volume} {63}},\ \bibinfo {pages} {279} (\bibinfo {year}
  {2015})}\BibitemShut {NoStop}%
\bibitem [{\citenamefont {Gu}(2010)}]{Gu2010}%
  \BibitemOpen
  \bibfield  {author} {\bibinfo {author} {\bibfnamefont {S.-J.}\ \bibnamefont
  {Gu}},\ }\href@noop {} {\bibfield  {journal} {\bibinfo  {journal} {Int. J.
  Modern Phys. B}\ }\textbf {\bibinfo {volume} {24}},\ \bibinfo {pages} {4371}
  (\bibinfo {year} {2010})}\BibitemShut {NoStop}%
\bibitem [{\citenamefont {Verstraete}\ \emph {et~al.}(2004)\citenamefont
  {Verstraete}, \citenamefont {Popp},\ and\ \citenamefont
  {Cirac}}]{Verstraete2004c}%
  \BibitemOpen
  \bibfield  {author} {\bibinfo {author} {\bibfnamefont {F.}~\bibnamefont
  {Verstraete}}, \bibinfo {author} {\bibfnamefont {M.}~\bibnamefont {Popp}}, \
  and\ \bibinfo {author} {\bibfnamefont {J.~I.}\ \bibnamefont {Cirac}},\
  }\href@noop {} {\bibfield  {journal} {\bibinfo  {journal} {Phys. Rev. Lett.}\
  }\textbf {\bibinfo {volume} {92}},\ \bibinfo {pages} {027901} (\bibinfo
  {year} {2004})}\BibitemShut {NoStop}%
\bibitem [{\citenamefont {Braunstein}\ and\ \citenamefont
  {Caves}(1994)}]{BraunsteinCaves1994}%
  \BibitemOpen
  \bibfield  {author} {\bibinfo {author} {\bibfnamefont {S.~L.}\ \bibnamefont
  {Braunstein}}\ and\ \bibinfo {author} {\bibfnamefont {C.~M.}\ \bibnamefont
  {Caves}},\ }\href@noop {} {\bibfield  {journal} {\bibinfo  {journal} {Phys.
  Rev. Lett.}\ }\textbf {\bibinfo {volume} {72}},\ \bibinfo {pages} {3439}
  (\bibinfo {year} {1994})}\BibitemShut {NoStop}%
\bibitem [{Note1()}]{Note1}%
  \BibitemOpen
  \bibinfo {note} {Interestingly, similar correlations, but in imaginary time,
  are being used to make the fidelity susceptibility calculable in quantum
  Monte Carlo computations. This computational technique also bounds the QFI
  (see Ref.~\cite{LeiWang2015} and references therein). Here, by relating the
  QFI to response functions in real time, we obtain a tool to measure it
  directly in laboratory experiments.}\BibitemShut {Stop}%
\bibitem [{sup()}]{supmatQFI}%
  \BibitemOpen
  \href@noop {} {\bibinfo  {journal} {See Supplementary Material}\
  }\BibitemShut {NoStop}%
\bibitem [{\citenamefont {St{\"o}ferle}\ \emph {et~al.}(2004)\citenamefont
  {St{\"o}ferle}, \citenamefont {Moritz}, \citenamefont {Schori}, \citenamefont
  {K{\"o}hl},\ and\ \citenamefont {Esslinger}}]{Stoferle2004}%
  \BibitemOpen
\bibfield  {journal} {  }\bibfield  {author} {\bibinfo {author} {\bibfnamefont
  {T.}~\bibnamefont {St{\"o}ferle}}, \bibinfo {author} {\bibfnamefont
  {H.}~\bibnamefont {Moritz}}, \bibinfo {author} {\bibfnamefont
  {C.}~\bibnamefont {Schori}}, \bibinfo {author} {\bibfnamefont
  {M.}~\bibnamefont {K{\"o}hl}}, \ and\ \bibinfo {author} {\bibfnamefont
  {T.}~\bibnamefont {Esslinger}},\ }\href@noop {} {\bibfield  {journal}
  {\bibinfo  {journal} {Phys. Rev. Lett.}\ }\textbf {\bibinfo {volume} {92}},\
  \bibinfo {pages} {130403} (\bibinfo {year} {2004})}\BibitemShut {NoStop}%
\bibitem [{\citenamefont {Ernst}\ \emph {et~al.}(2010)\citenamefont {Ernst},
  \citenamefont {G{\"o}tze}, \citenamefont {Krauser}, \citenamefont {Pyka},
  \citenamefont {L{\"u}hmann}, \citenamefont {Pfannkuche},\ and\ \citenamefont
  {Sengstock}}]{Ernst2010}%
  \BibitemOpen
  \bibfield  {author} {\bibinfo {author} {\bibfnamefont {P.~T.}\ \bibnamefont
  {Ernst}}, \bibinfo {author} {\bibfnamefont {S.}~\bibnamefont {G{\"o}tze}},
  \bibinfo {author} {\bibfnamefont {J.~S.}\ \bibnamefont {Krauser}}, \bibinfo
  {author} {\bibfnamefont {K.}~\bibnamefont {Pyka}}, \bibinfo {author}
  {\bibfnamefont {D.-S.}\ \bibnamefont {L{\"u}hmann}}, \bibinfo {author}
  {\bibfnamefont {D.}~\bibnamefont {Pfannkuche}}, \ and\ \bibinfo {author}
  {\bibfnamefont {K.}~\bibnamefont {Sengstock}},\ }\href@noop {} {\bibfield
  {journal} {\bibinfo  {journal} {Nature Phys.}\ }\textbf {\bibinfo {volume}
  {6}},\ \bibinfo {pages} {56} (\bibinfo {year} {2010})}\BibitemShut {NoStop}%
\bibitem [{\citenamefont {Parker}(2013)}]{Parker2013}%
  \BibitemOpen
  \bibfield  {author} {\bibinfo {author} {\bibfnamefont {S.~F.}\ \bibnamefont
  {Parker}},\ }\href@noop {} {\emph {\bibinfo {title} {Neutron scattering
  spectroscopy}}}\ (\bibinfo  {publisher} {Encyclopedia of Analytical Chemistry
  - Wiley Online Library},\ \bibinfo {year} {2013})\BibitemShut {NoStop}%
\bibitem [{\citenamefont {Sachdev}(1999)}]{Sachdev1999}%
  \BibitemOpen
  \bibfield  {author} {\bibinfo {author} {\bibfnamefont {S.}~\bibnamefont
  {Sachdev}},\ }\href@noop {} {\emph {\bibinfo {title} {Quantum Phase
  Transitions}}}\ (\bibinfo  {publisher} {Cambridge University Press},\
  \bibinfo {year} {1999})\BibitemShut {NoStop}%
\bibitem [{\citenamefont {H{\"a}lg}\ \emph {et~al.}(2015)\citenamefont
  {H{\"a}lg}, \citenamefont {H{\"u}vonen}, \citenamefont {Guidi}, \citenamefont
  {Quintero-Castro}, \citenamefont {Boehm}, \citenamefont {Regnault},
  \citenamefont {Hagiwara},\ and\ \citenamefont {Zheludev}}]{Haelg2015}%
  \BibitemOpen
  \bibfield  {author} {\bibinfo {author} {\bibfnamefont {M.}~\bibnamefont
  {H{\"a}lg}}, \bibinfo {author} {\bibfnamefont {D.}~\bibnamefont
  {H{\"u}vonen}}, \bibinfo {author} {\bibfnamefont {T.}~\bibnamefont {Guidi}},
  \bibinfo {author} {\bibfnamefont {D.~L.}\ \bibnamefont {Quintero-Castro}},
  \bibinfo {author} {\bibfnamefont {M.}~\bibnamefont {Boehm}}, \bibinfo
  {author} {\bibfnamefont {L.~P.}\ \bibnamefont {Regnault}}, \bibinfo {author}
  {\bibfnamefont {M.}~\bibnamefont {Hagiwara}}, \ and\ \bibinfo {author}
  {\bibfnamefont {A.}~\bibnamefont {Zheludev}},\ }\href@noop {} {\bibfield
  {journal} {\bibinfo  {journal} {Phys. Rev. B}\ }\textbf {\bibinfo {volume}
  {92}},\ \bibinfo {pages} {014412} (\bibinfo {year} {2015})}\BibitemShut
  {NoStop}%
\bibitem [{\citenamefont {T{\'o}th}(2005)}]{Toth2005}%
  \BibitemOpen
  \bibfield  {author} {\bibinfo {author} {\bibfnamefont {G.}~\bibnamefont
  {T{\'o}th}},\ }\href@noop {} {\bibfield  {journal} {\bibinfo  {journal}
  {Phys. Rev. A}\ }\textbf {\bibinfo {volume} {71}},\ \bibinfo {pages}
  {010301(R)} (\bibinfo {year} {2005})}\BibitemShut {NoStop}%
\bibitem [{\citenamefont {Wu}\ \emph {et~al.}(2005)\citenamefont {Wu},
  \citenamefont {Bandyopadhyay}, \citenamefont {Sarandy},\ and\ \citenamefont
  {Lidar}}]{Wu2005}%
  \BibitemOpen
  \bibfield  {author} {\bibinfo {author} {\bibfnamefont {L.-A.}\ \bibnamefont
  {Wu}}, \bibinfo {author} {\bibfnamefont {S.}~\bibnamefont {Bandyopadhyay}},
  \bibinfo {author} {\bibfnamefont {M.~S.}\ \bibnamefont {Sarandy}}, \ and\
  \bibinfo {author} {\bibfnamefont {D.~A.}\ \bibnamefont {Lidar}},\ }\href@noop
  {} {\bibfield  {journal} {\bibinfo  {journal} {Phys. Rev. A}\ }\textbf
  {\bibinfo {volume} {72}},\ \bibinfo {pages} {032309} (\bibinfo {year}
  {2005})}\BibitemShut {NoStop}%
\bibitem [{\citenamefont {Das}\ \emph {et~al.}(2006)\citenamefont {Das},
  \citenamefont {Sengupta}, \citenamefont {Sen},\ and\ \citenamefont
  {Chakrabarti}}]{Das2006}%
  \BibitemOpen
  \bibfield  {author} {\bibinfo {author} {\bibfnamefont {A.}~\bibnamefont
  {Das}}, \bibinfo {author} {\bibfnamefont {K.}~\bibnamefont {Sengupta}},
  \bibinfo {author} {\bibfnamefont {D.}~\bibnamefont {Sen}}, \ and\ \bibinfo
  {author} {\bibfnamefont {B.~K.}\ \bibnamefont {Chakrabarti}},\ }\href@noop {}
  {\bibfield  {journal} {\bibinfo  {journal} {Phys. Rev. B}\ }\textbf {\bibinfo
  {volume} {74}},\ \bibinfo {pages} {144423} (\bibinfo {year}
  {2006})}\BibitemShut {NoStop}%
\bibitem [{\citenamefont {Cardy}(1996)}]{bookCardy}%
  \BibitemOpen
  \bibfield  {author} {\bibinfo {author} {\bibfnamefont {J.}~\bibnamefont
  {Cardy}},\ }\href@noop {} {\emph {\bibinfo {title} {Scaling and
  Renormalization in Statistical Physics}}}\ (\bibinfo  {publisher} {Cambridge
  Lecture Notes in Physics},\ \bibinfo {year} {1996})\BibitemShut {NoStop}%
\bibitem [{\citenamefont {Jensen}\ and\ \citenamefont
  {Mackintosh}(1991)}]{JensenBookChapter3}%
  \BibitemOpen
  \bibfield  {author} {\bibinfo {author} {\bibfnamefont {J.}~\bibnamefont
  {Jensen}}\ and\ \bibinfo {author} {\bibfnamefont {A.~R.}\ \bibnamefont
  {Mackintosh}},\ }\enquote {\bibinfo {title} {Rare earth magnetism: Structures
  and excitations},}\ \ (\bibinfo  {publisher} {Clarendon Press, Oxford},\
  \bibinfo {year} {1991})\ Chap.\ \bibinfo {chapter} {Linear Response
  Theory}\BibitemShut {NoStop}%
\bibitem [{\citenamefont {Prokopenko}\ \emph {et~al.}(2011)\citenamefont
  {Prokopenko}, \citenamefont {Lizier}, \citenamefont {Obst},\ and\
  \citenamefont {Wang}}]{ProkopenkoObst2011}%
  \BibitemOpen
  \bibfield  {author} {\bibinfo {author} {\bibfnamefont {M.}~\bibnamefont
  {Prokopenko}}, \bibinfo {author} {\bibfnamefont {J.~T.}\ \bibnamefont
  {Lizier}}, \bibinfo {author} {\bibfnamefont {O.}~\bibnamefont {Obst}}, \ and\
  \bibinfo {author} {\bibfnamefont {X.~R.}\ \bibnamefont {Wang}},\ }\href@noop
  {} {\bibfield  {journal} {\bibinfo  {journal} {Phys. Rev. E}\ }\textbf
  {\bibinfo {volume} {84}},\ \bibinfo {pages} {041116} (\bibinfo {year}
  {2011})}\BibitemShut {NoStop}%
\bibitem [{\citenamefont {Greif}\ \emph {et~al.}(2013)\citenamefont {Greif},
  \citenamefont {Uehlinger}, \citenamefont {Jotzu}, \citenamefont {Tarruell},\
  and\ \citenamefont {Esslinger}}]{Greif2013}%
  \BibitemOpen
  \bibfield  {author} {\bibinfo {author} {\bibfnamefont {D.}~\bibnamefont
  {Greif}}, \bibinfo {author} {\bibfnamefont {T.}~\bibnamefont {Uehlinger}},
  \bibinfo {author} {\bibfnamefont {G.}~\bibnamefont {Jotzu}}, \bibinfo
  {author} {\bibfnamefont {L.}~\bibnamefont {Tarruell}}, \ and\ \bibinfo
  {author} {\bibfnamefont {T.}~\bibnamefont {Esslinger}},\ }\href@noop {}
  {\bibfield  {journal} {\bibinfo  {journal} {Science}\ }\textbf {\bibinfo
  {volume} {340}},\ \bibinfo {pages} {1307} (\bibinfo {year}
  {2013})}\BibitemShut {NoStop}%
\bibitem [{\citenamefont {Hart}\ \emph {et~al.}(2015)\citenamefont {Hart},
  \citenamefont {Duarte}, \citenamefont {Yang}, \citenamefont {Liu},
  \citenamefont {Paiva}, \citenamefont {Khatami}, \citenamefont {Scalettar},
  \citenamefont {Trivedi}, \citenamefont {Huse},\ and\ \citenamefont
  {Hulet}}]{Hart2015}%
  \BibitemOpen
  \bibfield  {author} {\bibinfo {author} {\bibfnamefont {R.~A.}\ \bibnamefont
  {Hart}}, \bibinfo {author} {\bibfnamefont {P.~M.}\ \bibnamefont {Duarte}},
  \bibinfo {author} {\bibfnamefont {T.-L.}\ \bibnamefont {Yang}}, \bibinfo
  {author} {\bibfnamefont {X.}~\bibnamefont {Liu}}, \bibinfo {author}
  {\bibfnamefont {T.}~\bibnamefont {Paiva}}, \bibinfo {author} {\bibfnamefont
  {E.}~\bibnamefont {Khatami}}, \bibinfo {author} {\bibfnamefont {R.~T.}\
  \bibnamefont {Scalettar}}, \bibinfo {author} {\bibfnamefont {N.}~\bibnamefont
  {Trivedi}}, \bibinfo {author} {\bibfnamefont {D.~A.}\ \bibnamefont {Huse}}, \
  and\ \bibinfo {author} {\bibfnamefont {R.~G.}\ \bibnamefont {Hulet}},\
  }\href@noop {} {\bibfield  {journal} {\bibinfo  {journal} {Nature}\ }\textbf
  {\bibinfo {volume} {519}},\ \bibinfo {pages} {211} (\bibinfo {year}
  {2015})}\BibitemShut {NoStop}%
\bibitem [{\citenamefont {Amico}\ \emph {et~al.}(2008)\citenamefont {Amico},
  \citenamefont {Fazio}, \citenamefont {Osterloh},\ and\ \citenamefont
  {Vedral}}]{Amico2008}%
  \BibitemOpen
  \bibfield  {author} {\bibinfo {author} {\bibfnamefont {L.}~\bibnamefont
  {Amico}}, \bibinfo {author} {\bibfnamefont {R.}~\bibnamefont {Fazio}},
  \bibinfo {author} {\bibfnamefont {A.}~\bibnamefont {Osterloh}}, \ and\
  \bibinfo {author} {\bibfnamefont {V.}~\bibnamefont {Vedral}},\ }\href@noop {}
  {\bibfield  {journal} {\bibinfo  {journal} {Rev. Mod. Phys.}\ }\textbf
  {\bibinfo {volume} {80}},\ \bibinfo {pages} {517} (\bibinfo {year}
  {2008})}\BibitemShut {NoStop}%
\bibitem [{\citenamefont {G{\"u}hne}\ and\ \citenamefont
  {T{\'o}th}(2009)}]{Guehne2009}%
  \BibitemOpen
  \bibfield  {author} {\bibinfo {author} {\bibfnamefont {O.}~\bibnamefont
  {G{\"u}hne}}\ and\ \bibinfo {author} {\bibfnamefont {G.}~\bibnamefont
  {T{\'o}th}},\ }\href@noop {} {\bibfield  {journal} {\bibinfo  {journal}
  {Physics Reports}\ }\textbf {\bibinfo {volume} {474}},\ \bibinfo {pages}
  {175} (\bibinfo {year} {2009})}\BibitemShut {NoStop}%
\bibitem [{\citenamefont {Ghosh}\ \emph {et~al.}(2003)\citenamefont {Ghosh},
  \citenamefont {Rosenbaum}, \citenamefont {Aeppli},\ and\ \citenamefont
  {Coppersmith}}]{Ghosh2003}%
  \BibitemOpen
  \bibfield  {author} {\bibinfo {author} {\bibfnamefont {S.}~\bibnamefont
  {Ghosh}}, \bibinfo {author} {\bibfnamefont {T.~F.}\ \bibnamefont
  {Rosenbaum}}, \bibinfo {author} {\bibfnamefont {G.}~\bibnamefont {Aeppli}}, \
  and\ \bibinfo {author} {\bibfnamefont {S.~N.}\ \bibnamefont {Coppersmith}},\
  }\href@noop {} {\bibfield  {journal} {\bibinfo  {journal} {Nature}\ }\textbf
  {\bibinfo {volume} {425}},\ \bibinfo {pages} {48} (\bibinfo {year}
  {2003})}\BibitemShut {NoStop}%
\bibitem [{\citenamefont {Brukner}\ \emph {et~al.}(2006)\citenamefont
  {Brukner}, \citenamefont {Vedral},\ and\ \citenamefont
  {Zeilinger}}]{Brukner2006}%
  \BibitemOpen
  \bibfield  {author} {\bibinfo {author} {\bibfnamefont {{\v C}.}~\bibnamefont
  {Brukner}}, \bibinfo {author} {\bibfnamefont {V.}~\bibnamefont {Vedral}}, \
  and\ \bibinfo {author} {\bibfnamefont {A.}~\bibnamefont {Zeilinger}},\
  }\href@noop {} {\bibfield  {journal} {\bibinfo  {journal} {Phys. Rev. A}\
  }\textbf {\bibinfo {volume} {73}},\ \bibinfo {pages} {012110} (\bibinfo
  {year} {2006})}\BibitemShut {NoStop}%
\bibitem [{\citenamefont {V{\'e}rtesi}\ and\ \citenamefont
  {Bene}(2006)}]{Vertesi2006}%
  \BibitemOpen
  \bibfield  {author} {\bibinfo {author} {\bibfnamefont {T.}~\bibnamefont
  {V{\'e}rtesi}}\ and\ \bibinfo {author} {\bibfnamefont {E.}~\bibnamefont
  {Bene}},\ }\href@noop {} {\bibfield  {journal} {\bibinfo  {journal} {Phys.
  Rev. B}\ }\textbf {\bibinfo {volume} {73}},\ \bibinfo {pages} {134404}
  (\bibinfo {year} {2006})}\BibitemShut {NoStop}%
\bibitem [{\citenamefont {Cramer}\ \emph {et~al.}(2013)\citenamefont {Cramer},
  \citenamefont {Bernard}, \citenamefont {Fabbri}, \citenamefont {Fallani},
  \citenamefont {Fort}, \citenamefont {Rosi}, \citenamefont {Caruso},
  \citenamefont {Inguscio},\ and\ \citenamefont {Plenio}}]{Cramer2013}%
  \BibitemOpen
  \bibfield  {author} {\bibinfo {author} {\bibfnamefont {M.}~\bibnamefont
  {Cramer}}, \bibinfo {author} {\bibfnamefont {A.}~\bibnamefont {Bernard}},
  \bibinfo {author} {\bibfnamefont {N.}~\bibnamefont {Fabbri}}, \bibinfo
  {author} {\bibfnamefont {L.}~\bibnamefont {Fallani}}, \bibinfo {author}
  {\bibfnamefont {C.}~\bibnamefont {Fort}}, \bibinfo {author} {\bibfnamefont
  {S.}~\bibnamefont {Rosi}}, \bibinfo {author} {\bibfnamefont {F.}~\bibnamefont
  {Caruso}}, \bibinfo {author} {\bibfnamefont {M.}~\bibnamefont {Inguscio}}, \
  and\ \bibinfo {author} {\bibfnamefont {M.}~\bibnamefont {Plenio}},\
  }\href@noop {} {\bibfield  {journal} {\bibinfo  {journal} {Nat. Commun.}\
  }\textbf {\bibinfo {volume} {4}},\ \bibinfo {pages} {2161} (\bibinfo {year}
  {2013})}\BibitemShut {NoStop}%
\bibitem [{\citenamefont {Wimmer}(2012)}]{Wimmer2012}%
  \BibitemOpen
  \bibfield  {author} {\bibinfo {author} {\bibfnamefont {M.}~\bibnamefont
  {Wimmer}},\ }\href@noop {} {\bibfield  {journal} {\bibinfo  {journal} {ACM
  Trans. Math. Software}\ }\textbf {\bibinfo {volume} {38}},\ \bibinfo {pages}
  {30} (\bibinfo {year} {2012})}\BibitemShut {NoStop}%
\bibitem [{\citenamefont {Derzhko}\ and\ \citenamefont
  {Krokhmalskii}(1997)}]{Derzhko1997}%
  \BibitemOpen
  \bibfield  {author} {\bibinfo {author} {\bibfnamefont {O.}~\bibnamefont
  {Derzhko}}\ and\ \bibinfo {author} {\bibfnamefont {T.}~\bibnamefont
  {Krokhmalskii}},\ }\href@noop {} {\bibfield  {journal} {\bibinfo  {journal}
  {Phys. Rev. B}\ }\textbf {\bibinfo {volume} {56}},\ \bibinfo {pages} {11663}
  (\bibinfo {year} {1997})}\BibitemShut {NoStop}%
\bibitem [{\citenamefont {Paredes}\ \emph {et~al.}(2004)\citenamefont
  {Paredes}, \citenamefont {Widera}, \citenamefont {Murg}, \citenamefont
  {Mandel}, \citenamefont {F{\"o}lling}, \citenamefont {Cirac}, \citenamefont
  {Shlyapnikov}, \citenamefont {H{\"a}nsch},\ and\ \citenamefont
  {Bloch}}]{Paredes2004}%
  \BibitemOpen
  \bibfield  {author} {\bibinfo {author} {\bibfnamefont {B.}~\bibnamefont
  {Paredes}}, \bibinfo {author} {\bibfnamefont {A.}~\bibnamefont {Widera}},
  \bibinfo {author} {\bibfnamefont {V.}~\bibnamefont {Murg}}, \bibinfo {author}
  {\bibfnamefont {O.}~\bibnamefont {Mandel}}, \bibinfo {author} {\bibfnamefont
  {S.}~\bibnamefont {F{\"o}lling}}, \bibinfo {author} {\bibfnamefont {J.~I.}\
  \bibnamefont {Cirac}}, \bibinfo {author} {\bibfnamefont {G.~V.}\ \bibnamefont
  {Shlyapnikov}}, \bibinfo {author} {\bibfnamefont {T.~W.}\ \bibnamefont
  {H{\"a}nsch}}, \ and\ \bibinfo {author} {\bibfnamefont {I.}~\bibnamefont
  {Bloch}},\ }\href@noop {} {\bibfield  {journal} {\bibinfo  {journal}
  {Nature}\ }\textbf {\bibinfo {volume} {429}},\ \bibinfo {pages} {277}
  (\bibinfo {year} {2004})}\BibitemShut {NoStop}%
\bibitem [{\citenamefont {Giampaolo}\ \emph {et~al.}(2008)\citenamefont
  {Giampaolo}, \citenamefont {Adesso},\ and\ \citenamefont
  {Illuminati}}]{Giampaolo2008}%
  \BibitemOpen
  \bibfield  {author} {\bibinfo {author} {\bibfnamefont {S.~M.}\ \bibnamefont
  {Giampaolo}}, \bibinfo {author} {\bibfnamefont {G.}~\bibnamefont {Adesso}}, \
  and\ \bibinfo {author} {\bibfnamefont {F.}~\bibnamefont {Illuminati}},\
  }\href@noop {} {\bibfield  {journal} {\bibinfo  {journal} {Phys. Rev. Lett.}\
  }\textbf {\bibinfo {volume} {100}},\ \bibinfo {pages} {197201} (\bibinfo
  {year} {2008})}\BibitemShut {NoStop}%
\bibitem [{\citenamefont {Derzhko}\ \emph {et~al.}(2000)\citenamefont
  {Derzhko}, \citenamefont {Krokhmalskii},\ and\ \citenamefont
  {Stolze}}]{Derzhko2000}%
  \BibitemOpen
  \bibfield  {author} {\bibinfo {author} {\bibfnamefont {O.}~\bibnamefont
  {Derzhko}}, \bibinfo {author} {\bibfnamefont {T.}~\bibnamefont
  {Krokhmalskii}}, \ and\ \bibinfo {author} {\bibfnamefont {J.}~\bibnamefont
  {Stolze}},\ }\href@noop {} {\bibfield  {journal} {\bibinfo  {journal} {J.
  Phys. A: Math. Gen.}\ }\textbf {\bibinfo {volume} {33}},\ \bibinfo {pages}
  {3063} (\bibinfo {year} {2000})}\BibitemShut {NoStop}%
\bibitem [{\citenamefont {Roux}\ \emph {et~al.}(2013)\citenamefont {Roux},
  \citenamefont {Minguzzi},\ and\ \citenamefont {Roscilde}}]{Roux2013}%
  \BibitemOpen
  \bibfield  {author} {\bibinfo {author} {\bibfnamefont {G.}~\bibnamefont
  {Roux}}, \bibinfo {author} {\bibfnamefont {A.}~\bibnamefont {Minguzzi}}, \
  and\ \bibinfo {author} {\bibfnamefont {T.}~\bibnamefont {Roscilde}},\
  }\href@noop {} {\bibfield  {journal} {\bibinfo  {journal} {New J. Phys.}\
  }\textbf {\bibinfo {volume} {15}},\ \bibinfo {pages} {055003} (\bibinfo
  {year} {2013})}\BibitemShut {NoStop}%
\bibitem [{\citenamefont {G{\"u}hne}\ and\ \citenamefont
  {T{\'o}th}(2006)}]{Guehne2006}%
  \BibitemOpen
  \bibfield  {author} {\bibinfo {author} {\bibfnamefont {O.}~\bibnamefont
  {G{\"u}hne}}\ and\ \bibinfo {author} {\bibfnamefont {G.}~\bibnamefont
  {T{\'o}th}},\ }\href@noop {} {\bibfield  {journal} {\bibinfo  {journal}
  {Phys. Rev. A}\ }\textbf {\bibinfo {volume} {73}},\ \bibinfo {pages} {052319}
  (\bibinfo {year} {2006})}\BibitemShut {NoStop}%
\bibitem [{\citenamefont {Hide}\ \emph {et~al.}(2007)\citenamefont {Hide},
  \citenamefont {Son}, \citenamefont {Lawrie},\ and\ \citenamefont
  {Vedral}}]{Hide2007}%
  \BibitemOpen
  \bibfield  {author} {\bibinfo {author} {\bibfnamefont {J.}~\bibnamefont
  {Hide}}, \bibinfo {author} {\bibfnamefont {W.}~\bibnamefont {Son}}, \bibinfo
  {author} {\bibfnamefont {I.}~\bibnamefont {Lawrie}}, \ and\ \bibinfo {author}
  {\bibfnamefont {V.}~\bibnamefont {Vedral}},\ }\href@noop {} {\bibfield
  {journal} {\bibinfo  {journal} {Phys. Rev. A}\ }\textbf {\bibinfo {volume}
  {76}},\ \bibinfo {pages} {022319} (\bibinfo {year} {2007})}\BibitemShut
  {NoStop}%
\bibitem [{\citenamefont {Botet}\ and\ \citenamefont
  {Jullien}(1983)}]{Botet1983}%
  \BibitemOpen
  \bibfield  {author} {\bibinfo {author} {\bibfnamefont {R.}~\bibnamefont
  {Botet}}\ and\ \bibinfo {author} {\bibfnamefont {R.}~\bibnamefont
  {Jullien}},\ }\href@noop {} {\bibfield  {journal} {\bibinfo  {journal} {Phys.
  Rev. B}\ }\textbf {\bibinfo {volume} {28}},\ \bibinfo {pages} {3955}
  (\bibinfo {year} {1983})}\BibitemShut {NoStop}%
\bibitem [{\citenamefont {Thomas}(1925)}]{Thomas1925}%
  \BibitemOpen
  \bibfield  {author} {\bibinfo {author} {\bibfnamefont {W.}~\bibnamefont
  {Thomas}},\ }\href@noop {} {\bibfield  {journal} {\bibinfo  {journal}
  {Naturwiss.}\ }\textbf {\bibinfo {volume} {28}},\ \bibinfo {pages} {627}
  (\bibinfo {year} {1925})}\BibitemShut {NoStop}%
\bibitem [{\citenamefont {Reiche}\ and\ \citenamefont
  {Thomas}(1925)}]{Reiche1925}%
  \BibitemOpen
  \bibfield  {author} {\bibinfo {author} {\bibfnamefont {F.}~\bibnamefont
  {Reiche}}\ and\ \bibinfo {author} {\bibfnamefont {W.}~\bibnamefont
  {Thomas}},\ }\href@noop {} {\bibfield  {journal} {\bibinfo  {journal} {Z.
  Phys.}\ }\textbf {\bibinfo {volume} {34}},\ \bibinfo {pages} {510} (\bibinfo
  {year} {1925})}\BibitemShut {NoStop}%
\bibitem [{\citenamefont {Kuhn}(1925)}]{Kuhn1925}%
  \BibitemOpen
  \bibfield  {author} {\bibinfo {author} {\bibfnamefont {W.}~\bibnamefont
  {Kuhn}},\ }\href@noop {} {\bibfield  {journal} {\bibinfo  {journal} {Z.
  Phys.}\ }\textbf {\bibinfo {volume} {33}},\ \bibinfo {pages} {408} (\bibinfo
  {year} {1925})}\BibitemShut {NoStop}%
\bibitem [{\citenamefont {Lucarini}\ \emph {et~al.}(2003)\citenamefont
  {Lucarini}, \citenamefont {Bassani}, \citenamefont {Peiponen},\ and\
  \citenamefont {Saarinen}}]{Lucarini2003}%
  \BibitemOpen
  \bibfield  {author} {\bibinfo {author} {\bibfnamefont {V.}~\bibnamefont
  {Lucarini}}, \bibinfo {author} {\bibfnamefont {F.}~\bibnamefont {Bassani}},
  \bibinfo {author} {\bibfnamefont {K.-E.}\ \bibnamefont {Peiponen}}, \ and\
  \bibinfo {author} {\bibfnamefont {J.~J.}\ \bibnamefont {Saarinen}},\
  }\href@noop {} {\bibfield  {journal} {\bibinfo  {journal} {Rivista del Nuovo
  Cimento}\ }\textbf {\bibinfo {volume} {26}} (\bibinfo {year}
  {2003})}\BibitemShut {NoStop}%
 \bibitem [{\citenamefont {Wang}\ \emph {et~al.}(2015)\citenamefont {Wang},
  \citenamefont {Liu}, \citenamefont {Imri{\v s}ka}, \citenamefont {Ma},\ and\
  \citenamefont {Troyer}}]{LeiWang2015}%
  \BibitemOpen
  \bibfield  {author} {\bibinfo {author} {\bibfnamefont {L.}~\bibnamefont
  {Wang}}, \bibinfo {author} {\bibfnamefont {Y.-H.}\ \bibnamefont {Liu}},
  \bibinfo {author} {\bibfnamefont {J.}~\bibnamefont {Imri{\v s}ka}}, \bibinfo
  {author} {\bibfnamefont {P.~N.}\ \bibnamefont {Ma}}, \ and\ \bibinfo {author}
  {\bibfnamefont {M.}~\bibnamefont {Troyer}},\ }\href@noop {} {\bibfield
  {journal} {\bibinfo  {journal} {Phys. Rev. X}\ }\textbf {\bibinfo {volume}
  {5}},\ \bibinfo {pages} {031007} (\bibinfo {year} {2015})}\BibitemShut
  {NoStop}%
\end{thebibliography}
\end{document}